\documentclass[preprint,3p,12p]{elsarticle}

\usepackage{amssymb}
\usepackage{amsmath}
\usepackage{mathtools}
\usepackage{amsthm}
\usepackage{standalone}
\usepackage{booktabs}
\usepackage{layouts}
\usepackage{xparse,mathtools}
\usepackage{multirow}
\usepackage{array}
\usepackage{pdflscape}
\usepackage{arydshln}
\usepackage{multicol,multirow}
\usepackage{amsfonts}
\usepackage{soul}
\usepackage{rotating}
\usepackage{textcomp}
\usepackage{xcolor}
\usepackage[colorlinks,allcolors=blue]{hyperref}
\usepackage{enumitem}
\usepackage{layouts}
\usepackage[caption=false]{subfig}
\usepackage{microtype}
\newcommand\norm[1]{\left\lVert#1\right\rVert}

\begin{document}

\begin{frontmatter}

\title{Physics-informed, Generative Adversarial Design of Funicular Shells}

\author[1]{Rúben Lourenço}

\author[1]{Icíar Alfaro}

\author[2]{Beatriz Moya\corref{cor1}}
\ead{beatriz.moya_garcia@ensam.eu}
\author[1]{Elias Cueto}

\affiliation[1]{organization={Keysight-UZ Chair of the National Strategy on Artificial Intelligence. Aragon Institute of Engineering Research},
addressline={Universidad de Zaragoza}, 
city={Zaragoza},
citysep={},
postcode={50018}, 
country={Spain}}

\affiliation[2]{organization={PIMM lab},
addressline={Arts et M\'etiers Institute of Technology}, 
city={Paris},
citysep={}, 
postcode={75013}, 
country={France}}

\cortext[cor1]{Corresponding author at: PIMM lab, Arts et Métiers Institute of Technology, Bd de l'Hôpital 153, 75013 Paris, France.}

%%%%%%%%%%%%%%%%%%%%%%%%%%%%%%%%%%%%%%
%% Abstract
%%%%%%%%%%%%%%%%%%%%%%%%%%%%%%%%%%%%%%
\begin{abstract}
Shell structures are pivotal in the fields of architecture and engineering, due to their
aesthetic appeal and structural efficiency. Recently, 3D concrete printing has reignited the interest in these structures. But, as printed concrete cannot be reinforced with steel, structures built in this way must be designed to withstand primarily pure compression: they must be funicular shells. Nevertheless, a fundamental challenge remains unsolved since Robert Hooke’s discovered the catenary arch in 1675: it is not known whether the concept of a funicular polygon can be generalised to three-dimensional structures. 

Generative Adversarial Networks (GANs), have shown remarkable success in generating realistic data samples matching the distribution of the training data and have been shown to produce highly convincing synthetic images. This work proposes a physics-informed generative adversarial framework for the design of funicular shell structures. The approach employs a modified Deep Convolutional Generative Adversarial architecture physically guided by an auxiliary discriminator to generate realistic and structurally efficient shell geometries. Specifically, the model is constrained by the membrane factor to penalize geometries dominated by bending. An additional discriminator is also employed allowing the model to deal with more complex structures. Results show that the developed model is stable and capable of generating physically optimal, previously unseen, funicular shells with smooth forms and high membrane factor distributions.
\end{abstract}

%%Research highlights
\begin{highlights}
\item A physics-informed generative framework is proposed for the design of funicular shell structures.
\item A pre-trained PB-PUNet model serves as an auxiliary discriminator to guide generation based on the membrane factor.
\item The AD-DCGAN incorporates a mask discriminator and PCA-guided latent space to handle complex geometries and mitigate mode collapse.
\item The PB-PUNet surrogate model accurately predicts shell structural behaviour, achieving total errors as low as 2.2\% on the test set.
\item The AD-DCGAN generates diverse, structurally efficient shell geometries with high membrane factors, validated through finite element analysis.
\end{highlights}

%% Keywords

\end{frontmatter}

%%%%%%%%%%%%%%%%%%%%%%%%%%%%%%%%%%%%
%% Introduction
%%%%%%%%%%%%%%%%%%%%%%%%%%%%%%%%%%%%
\clearpage
\section{Introduction}

In 1675, Robert Hooke discovered what he called the ``true mathematical and mechanical form'' of arches in building construction. According to Hooke, an arch subjected to its own weight experiences only compressive forces when its shape corresponds to that of an inverted hanging chain \cite{hooke1676}. Later, Leibniz, Huygens, and Bernoulli derived the equation describing this geometry, which became known as the catenary arch \cite{hill1962}. The principle was later extended to arbitrary loading conditions, with the resulting equilibrium shape termed the funicular form \cite{block2006}. Since its formulation, this theory has been applied to a wide range of structural shapes \cite{block2006hangs}. However, despite extensive research, the identification of optimal three-dimensional shell forms based on the same fundamental criteria remains an open topic of research \cite{adriaenssens2014}.

Concrete shells assume a pivotal role in building design, architecture and engineering, being valued for their structural efficiency and aesthetic qualities. Nevertheless, their widespread adoption has declined in recent decades largely due to shifts in architectural movements and the complexity associated with traditional formwork-based construction methods. However, the advent of three-dimensional concrete printing (3DCP) has renewed the interest in funicular shell structures, offering the potential to overcome several limitations of the conventional concrete construction methods \cite{bos2016}. 3DCP mitigates the labour-intensive nature of traditional methods, reduces material waste and lowers the carbon footprint associated with bulk concrete usage. While 3DCP is limited to specific geometries dictated by the layer-by-layer deposition process, it also requires careful design to ensure that, throughout its service life, the structure is subjected only to compressive stresses, thereby avoiding tensile stresses, which it is unable to withstand given the practical impossibility of incorporating steel reinforcement \cite{mader2023,bhooshan2022striatus}.

The design of compression-only (funicular) shell structures  involves a complex interplay between structural stability, load distribution and material efficiency, striking the balance between mechanical, geometric, and aesthetic considerations. Given the practical significance of this problem, numerous computational approaches have been developed to support generative shell design, based on Finite Element Analysis (FEA), optimization algorithms or even Artificial Intelligence (AI) techniques \cite{adriaenssens2014}. In the experimental domain, Isler \cite{chilton2000} generated funicular shells by hanging plaster-impregnated fabrics under self-weight which, upon hardening, formed a rigid surface that was inverted to obtain the desired shape, replicating Hooke's hanging chain analogy. In the computational domain, most methods rely on the lower bound theorem of limit analysis, operating primarily with force equilibria rather than stress fields. Among these, Thrust Network Analysis has emerged as a prominent approach \cite{block2007}. Recently, Olivieri \textit{et al.} proposed an optimization-based method employing the Airy stress function, while Pastrana \textit{et al.} developed a framework combining neural networks with a differentiable mechanics simulator, adhering to the principles of the lower bound theorem \cite{olivieri2025, pastrana2025}. The latter approach not only yields physically admissible designs but also demonstrates improved computational efficiency relative to traditional optimization strategies.

Within the realm of artificial intelligence, generative adversarial networks (GANs) are a class of unsupervised learning models that have shown remarkable success in generating realistic data samples matching the statistical distribution of the training data \cite{goodfellow2014}. GANs have demonstrated their generative capability across a diverse range of domains, including image-to-image translation, three-dimensional object generation, data augmentation, and the synthesis of photorealistic imagery \cite{wu2017,li2025,smith2017}. Based on this success, the present study introduces an AI-driven methodology for the generative design of compression-only shell structures without assuming any type of simplification on their behaviour. A central objective is to circumvent optimization-based approaches that, while easily solvable via differentiable neural networks, are susceptible to converge to local minima. The primary aim is to develop a generative tool that assists structural designers in exploring efficient shell forms. To this end, a physics-informed generative framework based on a modified GAN architecture is proposed, capable of generating shell geometries that are both realistic and structurally efficient.

Several variants of physics-informed GANs have been reported in the literature. Chen \textit{et al.} introduced a physics-informed GAN (PI-GAN) incorporating a transition function within the generator to address the problem of sequence generation with limited data \cite{chen2020}. Daw \textit{et al.} proposed a physics-informed discriminator GAN (PID-GAN) for uncertainty quantification, wherein physics consistency scores serve as auxiliary inputs to the discriminator \cite{daw2021}. Ciftci and Hackl  developed a GAN-based framework for model-free data-driven computational mechanics, employing a physics-informed neural network (PINN)  as the generator \cite{ciftci2024,raissi2019,bastek2023}. Li \textit{et al.}  extended this framework to solve forward and inverse problems in low-data regimes \cite{li2025}. In contrast, the approach presented herein integrates a GAN with a pre-trained PINN that works as an auxiliary discriminator to ensure the physical compliance of generated shell structures, based on the membrane factor. The idea is inspired on the concept of dual discriminator GANs previously explored by Nguyen \textit{et al}., although those implementations primarily addressed training stability rather than enforcing physical consistency \cite{nguyen2017}. The PINN is pre-trained as a surrogate model to predict the spatial distribution of the membrane factor across the shell domain, ensuring an accurate physical evaluation prior to coupling with the GAN. Another core innovation of this approach is the inclusion of a third discriminator promoting the generation of binary masks that allow the GAN to deal with more complex geometries containing empty areas. %We end with the conclusions of this work in Section \ref{sc:conclusions}.

The paper is organized as follows. For completeness, Section \ref{sc:shell_theory} presents a brief overview of the theoretical foundations governing the mechanics of shell structures. Section \ref{sc:gans} reviews GANs and their mathematical formulation, including the techniques used to enhance training stability. Section \ref{sc:methods} details the proposed model architectures, data generation process, geometry representation, the training and validation procedures. Section \ref{sc:results} presents the results of a parametric study on the optimisation of the PINN model, as well as the predictions obtained using the optimised model based on test data. A set of generated surfaces is also presented, together with their application in FEA, to assess the quality of the GAN model. Finally, the paper closes in Section \ref{sc:conclusions} with the concluding remarks of this work.

%%%%%%%%%%%%%%%%%%%%%%%%%%%%%%%%%%%%
%% Shell theory
%%%%%%%%%%%%%%%%%%%%%%%%%%%%%%%%%%%%
\section{Assessment of the behaviour of the shells}
\label{sc:shell_theory}

In what follows, we assume that the shells of interest can be adequately described using the well-known Kirchhoff–Love theory. We will not repeat the theory in question here, as it is a classic and well known to the vast majority of readers. In general, it is described in detail in references such as \cite{reddy1999theory}. In thin, Kirchhoff-Love shells, the ratio between thickness and the shorter span length should be less than 1/20 \cite{ugural2009}. The shells can then be approximated by assuming that they form a 3D differential manifold, capturing the out-of-plane kinematics by some simplifying assumptions, which are valid as long as the thickness is small compared to the in-plane dimensions. The basic kinematic assumptions associated with the deformation of a thin shell under the scope of small-deformation analysis are \cite{ventsel2002,reddy2006}:
\begin{enumerate}[leftmargin=1.3\parindent, label=(\roman*)]
    \item the ratio of shell thickness to radius of curvature of the midsurface is small;
    \item deflections are small compared with shell thickness;
    \item straight lines normal to the midsurface remain straight and normal to the deformed midsurface after bending.
\end{enumerate}

The equilibrium equations for the shell are derived from the principle of minimum total
potential energy. The total potential energy $\mathcal{P}$ is the sum of the elastic strain energy $\mathcal{U}$ stored in the deformed shell and the potential energy $\mathcal{V}$ associated to the applied forces. When the shell is at equilibrium, the total potential energy is stationary and the first variation of the total potential energy remains zero, leading to the following equilibrium:
\begin{equation}
\delta\mathcal{P}=\delta\mathcal{U}+\delta\mathcal{V}=0.
\end{equation}
Assuming quasi-static loading conditions, no kinetic energy is present and the variation of the total potential energy of the shell can be expressed such that:
\begin{equation}
\label{eq:balance}\delta\mathcal{P}=\mathcal{W}_{\text{memb}}+\mathcal{W}_{\text{flex}}-\mathcal{W}_{\text{ext}} = 0,~
\end{equation}
where the membrane work $\mathcal{W}_{\text{memb}}$, the flexural work $\mathcal{W}_{\text{flex}}$, and the external work $\mathcal{W}_{\text{ext}}$ are defined such that:
\begin{align}
    \mathcal{W}_{\text{memb}}&=\int_\Gamma F_{11}\varepsilon_{11}^m+F_{22}\varepsilon_{22}^m+F_{12}\gamma_{12}^m~{d}\xi_1{d}\xi_2,~\label{eq:w_memb}\\
    \mathcal{W}_{\text{flex}}&=\int_\Gamma M_{11}\varepsilon_{11}^f+M_{22}\varepsilon_{22}^f+M_{12}\gamma_{12}^f~{d}\xi_1{d}\xi_2,~\label{eq:w_flex}\\
    \mathcal{W}_{\text{ext}}&=\int_\Gamma \boldsymbol{{f}}_{\text{ext}}^{\top}\cdot \boldsymbol{{u}}~{d}\xi_1{d}\xi_2,
\end{align}
where $\boldsymbol{\varepsilon}^m=[\varepsilon_{11}^m,\varepsilon_{22}^m,\varepsilon_{12}^m]$ and $\boldsymbol{\varepsilon}^f=[\varepsilon_{11}^f,\varepsilon_{22}^f,\varepsilon_{12}^f]$ are the membrane and flexural strains, while $\boldsymbol{{F}}=[F_{11}, F_{22}, F_{12}]$ and $\boldsymbol{{M}}=[M_{11}, M_{22}, M_{12}]$ are the stress resultants and the bending moments, respectively. Finally, $\boldsymbol{{f}}_{\text{ext}}$ represents the vector of externally applied loads, while $\boldsymbol{{u}}$ represents the displacement field.

With the variables obtained from Eqs. (\ref{eq:w_memb}) and (\ref{eq:w_flex}) the degree to which the mechanical behaviour of the shell is dominated either by membrane or bending effects can be evaluated through its membrane factor distribution ${{m}}_{{f}}$ defined as:
\begin{equation}
    {{m}}_{{f}}=\cfrac{\mathcal{W}_{\text{memb}}}{\mathcal{W}_{\text{memb}}+\mathcal{W}_{\text{flex}}}~,~m_{{f}}\in [0,1]~.
\end{equation}
For a membrane-dominated structure ${{m}}_{{f}}$ should be close to $1.0$, while for a bending-dominated structure ${{m}}_{{f}}$ will be close to $0.0$. For the case of 3D-printed concrete structures, the interest lies on obtaining structures where the membrane behaviour dominates, thus maximising the membrane factor. This eliminates the need for steel reinforcements, which cannot be incorporated due to the specific nature of the printing process.

%%%%%%%%%%%%%%%%%%%%%%%%%%%%%%%%%%%%
%% GANS
%%%%%%%%%%%%%%%%%%%%%%%%%%%%%%%%%%%%
\section{Generative adversarial networks}\label{sc:gans}

\subsection{Theoretical foundations}\label{ssc:gan_theory}

Supervised Machine Learning (ML) consists on finding a mapping between pairs of known input and output data. In
contrast, unsupervised learning extracts patterns from input data. Generative modelling in ML is the process of automatically learning the patterns in some input data set. As a result, the model is able to create new samples that could have been derived from the original data set. Generative adversarial networks (GANs) phrase this process in a supervised manner to foster the quality of the new samples, by combining two networks \cite{goodfellow2014}. The goal is to train a generator network $G(\boldsymbol{{z}};\theta_{{g}})$ to learn a distribution $p_{{g}}(\boldsymbol{{x}})$ from the original data distribution $p_{\text{data}}(\boldsymbol{{x}})$ through the transformation of a random noise vector $\boldsymbol{{z}}$. Simultaneously, a discriminator network $D(\boldsymbol{{x}};\theta_{{d}})$ is trained to classify the samples between fake or real. The generator is trained to fool the discriminator into accepting its outputs as being real, resulting in the following two-player minmax game \cite{goodfellow2014,salimans2016}: 
\begin{equation}
    \min_G\max_DV(D,G)=\mathbb{E}_{\boldsymbol{{x}}\sim p_{\text{data}}(\boldsymbol{{x}})}\log\left[D(\boldsymbol{{x}})\right]+\mathbb{E}_{\boldsymbol{{z}}\sim p_{{z}}(\boldsymbol{{z}})}\left[\log(1-D(G(\boldsymbol{{z}})))\right].
\end{equation}
When the generator distribution converges to the true data distribution $p_{{g}}(\boldsymbol{{x}})\approx p_{data}(\boldsymbol{{x}})$, the training processe reaches the Nash equilibrium and the discriminator converges to \cite{goodfellow2014}: 
\begin{equation}
    D(\boldsymbol{{x}})=\cfrac{p_{\text{data}}(\boldsymbol{{x}})}{p_{\text{data}}(\boldsymbol{{{x}}})+p_g(\boldsymbol{{x}})}=\cfrac{1}{2}~.
\end{equation}

\subsection{Deep convolutional GANs}

\begin{figure}[h!]%
\centering
\includegraphics[width=0.75\textwidth]{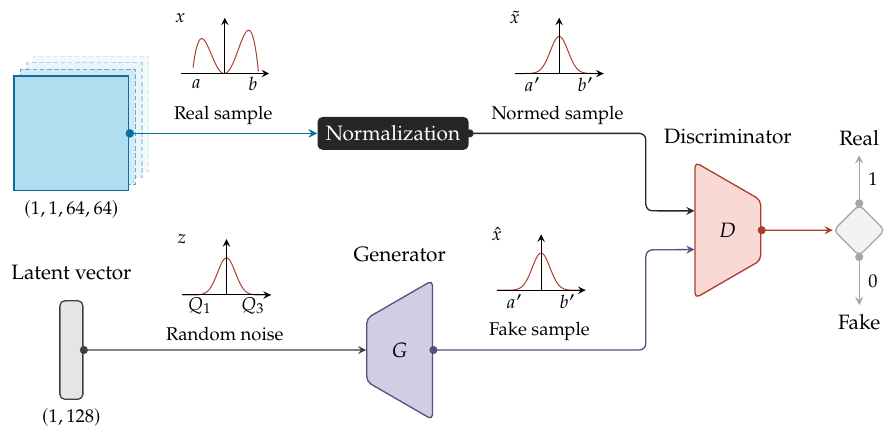}
\caption{Deep convolutional GAN architecture.}
\label{fig:dcgan}
\end{figure}

Generative models have shown a great success in generating high-quality synthetic data. In the seminal paper by Goodfellow \textit{et al.} \cite{goodfellow2014} the formulations presented in Section \ref{ssc:gan_theory} apply to dense multi-layer feedforward neural networks. However, GAN architectures are often implemented using Deep Convolutional GANs (DCGANs) to generate high-resolution images (see Fig. \ref{fig:dcgan}). Originally developed by Radford \textit{et al.} \cite{radford2016}, DCGANs employ convolutional layers in both the generator and discriminator. However, for the purpose of generative tasks, DCGANs introduced some modifications to the standard Convolutional Neural Network (CNN) architectures to ensure improved training stability. For a deeper insight at these modifications, the reader is invited to refer to the original paper \cite{radford2016}.

\subsection{Addressing mode collapse}

Due to the adversarial training dynamics of GANs and the non-convex nature of the loss landscape, reaching the Nash equilibrium is particularly challenging \cite{goodfellow2014}. This often leads to mode collapse, a common failure mode of GANs in which the generator collapses to a single mode of the data distribution, while the discriminator fails to distinguish between real and fake samples \cite{salimans2016, nguyen2017}. Over the years, numerous techniques have been proposed to mitigate mode collapse, see \cite{nguyen2017,salimans2016,miyato2018,arjovsky2017}. However, in this work two specific techniques are explicitly used to mitigate that issue: feature matching and spectral normalization.

\subsubsection{Feature matching}\label{ssc:feat_match}
Feature matching (FM) addresses training instability by introducing a regularized objective that prevents the generator from overtraining relative to the current discriminator state. With this new objective, the generator is trained to match the features $\boldsymbol{{d}}(\boldsymbol{{x}})$ of an intermediate layer of the discriminator. Specifically, the discriminator is tasked with identifying features that are most discriminative of real data, while the generator learns to produce samples whose feature representations align with those of real data. The feature matching objective for the generator is defined as \cite{salimans2016}:
\begin{equation}
\mathcal{L}_{\mathrm{FM}}=\norm{\mathbb{E}_{\boldsymbol{{x}}\sim p_{\text{data}}}\boldsymbol{{d}}\left(\boldsymbol{{x}}\right)-\mathbb{E}_{\boldsymbol{{z}}\sim p_{{z}}(\boldsymbol{{z}})}\boldsymbol{{d}}\left(G\left(\boldsymbol{{z}}\right)\right)}^2_2~.
\end{equation}
This formulation provides a more stable training signal than the original GAN objective, as the feature statistics evolve more gradually than the discriminator's binary classification output. 

\subsubsection{Spectral normalization}

Spectral normalization (SN) is a weight normalization technique that stabilizes GAN training by constraining the Lipschitz constant of the discriminator function \cite{miyato2018}. For a discriminator $D_{\theta_{{d}}}$, the Lipschitz norm $\lVert D_{\theta_{{d}}} \rVert_{\text{Lip}}$ is determined by the largest singular value $\sigma(\boldsymbol{{W}}_{{d}})$ of each weight matrix $\boldsymbol{{W}}_{{d}}$. Spectral normalization enforces Lipschitz continuity by normalizing each weight matrix as follows, see \cite{miyato2018}:
\begin{equation}
\boldsymbol{ \bar{W}}_{{d}_{\text{SN}}} = \frac{\boldsymbol{{W}}_{{d}}}{\sigma(\boldsymbol{{W}}_{{d}})},
\end{equation}
which ensures that $\lVert D_{\theta_{{d}}} \rVert_{\text{Lip}} \leq 1$. This Lipschitz constraint simultaneously addresses two fundamental failure modes in GAN training. First, it provides an upper bound on the discriminator's gradients, preventing explosive gradients. Second, by enforcing a structured variance across the discriminator's parameters, it mitigates the vanishing gradient problem that often leaves the generator without meaningful feedback \cite{miyato2018, lin2021}. The resulting smoother discriminator function provides more informative gradients to the generator, encouraging it to cover the full support of the data distribution rather than collapsing to isolated modes.

%%%%%%%%%%%%%%%%%%%%%%%%%%%%%%%%%%%%
%% Methodology
%%%%%%%%%%%%%%%%%%%%%%%%%%%%%%%%%%%%
\section{Materials and methods}\label{sc:methods}
%%%%%%%%%%%%%%%%%%%%%%%%%%%%%%%%%%%%
%% Data generation
%%%%%%%%%%%%%%%%%%%%%%%%%%%%%%%%%%%%
\subsection{Training database}\label{ssc:database}

To obtain a dataset large enough to train the GAN model, a database was created containing a
total of 6,058 samples. For the purpose, a Python-based procedural tool was developed to automate the geometry generation process, replicating Hooke's analogy, using the software Blender \cite{blender}. Each sample originated from a thin $10\times10$ m\textsuperscript{2} square plane, fixed at its corners, that could contain a central hole or not. A physics simulation was then performed in which the flat plane was assigned thin cloth properties, thus with a very low bending stiffness, and allowed to deform under gravitational load. Inspired by the work of Heinz Isler, the geometry is treated as fabric through a 3D mesh where the vertices are point masses with the edges between them acting as springs that resist stretching, shearing, and bending \cite{provot_a,provot_b,mezger}.

The plates used to model our structure do not need to be strictly cable-supported, as will be seen later. The method itself will determine the associated membrane factor, aided by a finite element analysis.

\begin{figure}[!htbp]
    \centering
    \subfloat[]{
        \includegraphics[width=0.22\textwidth]{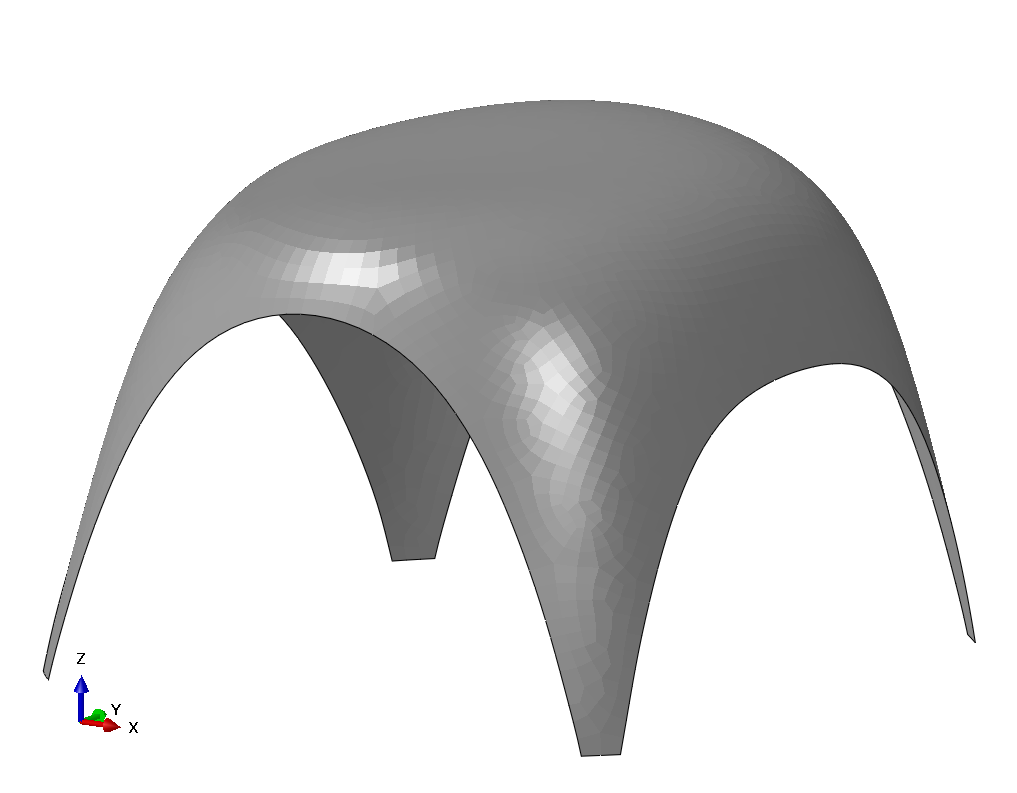}
    }\hspace{6mm}
    \subfloat[]{
        \includegraphics[width=0.22\textwidth]{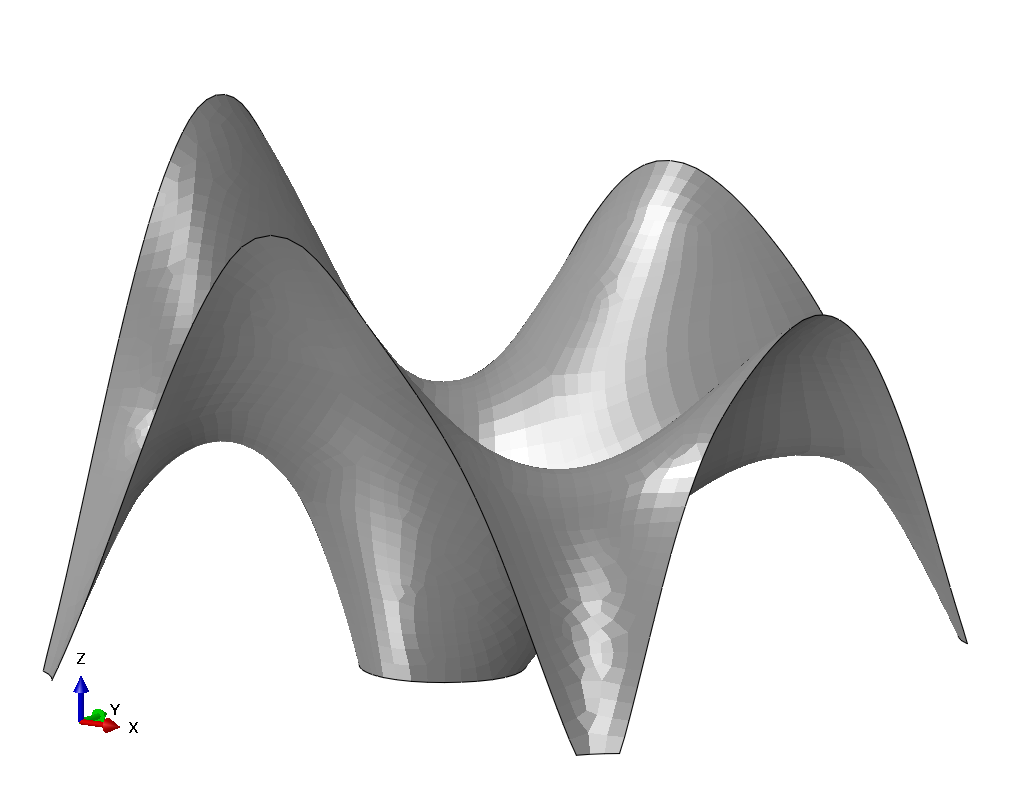}
    }\hspace{6mm}
    \subfloat[]{
        \includegraphics[width=0.22\textwidth]{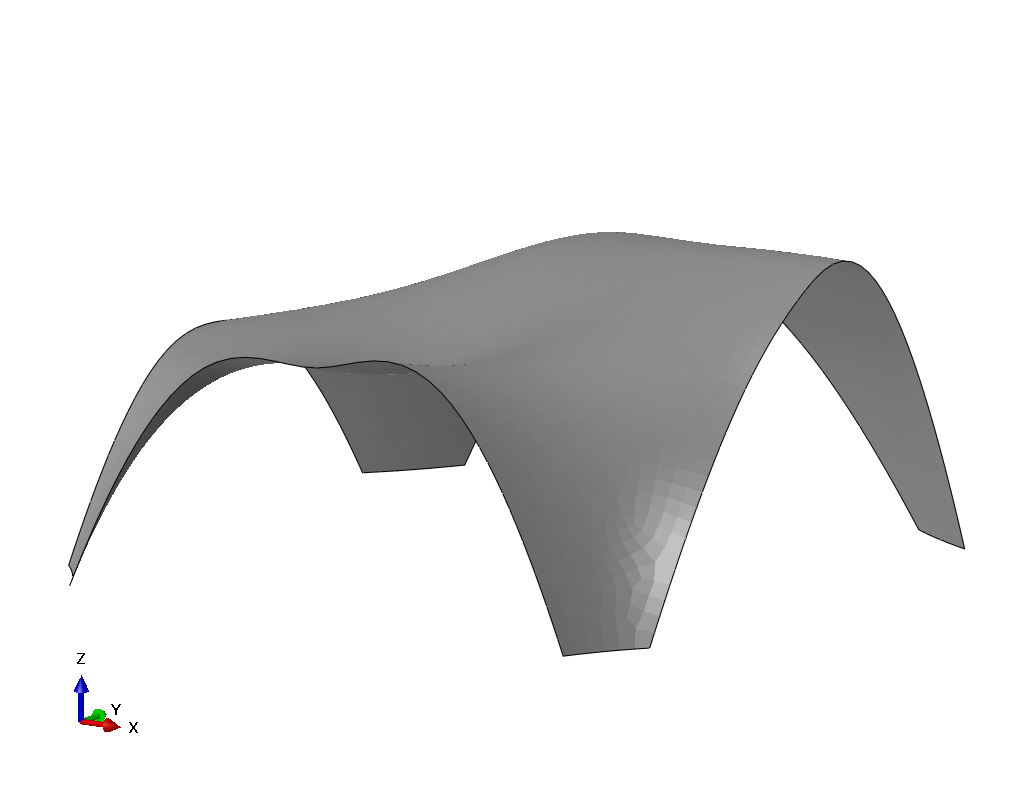}
    }\\[-1.55em]
    \subfloat[]{
        \includegraphics[width=0.22\textwidth]{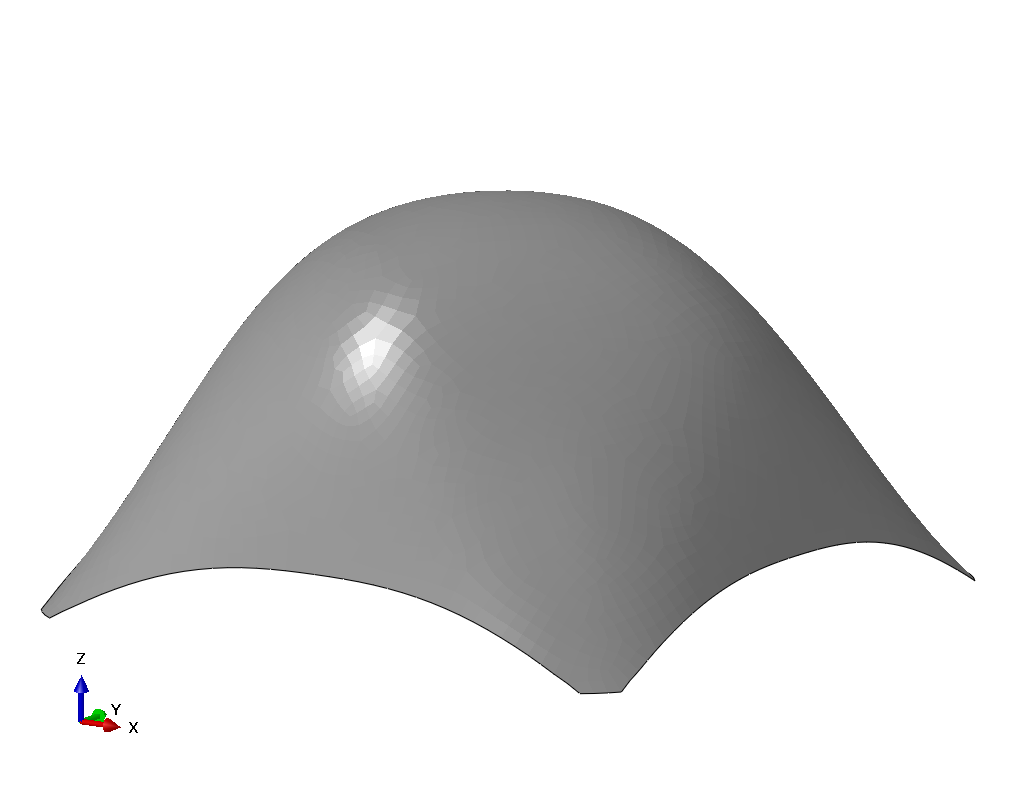}
    }\hspace{6mm}
    \subfloat[]{
        \includegraphics[width=0.22\textwidth]{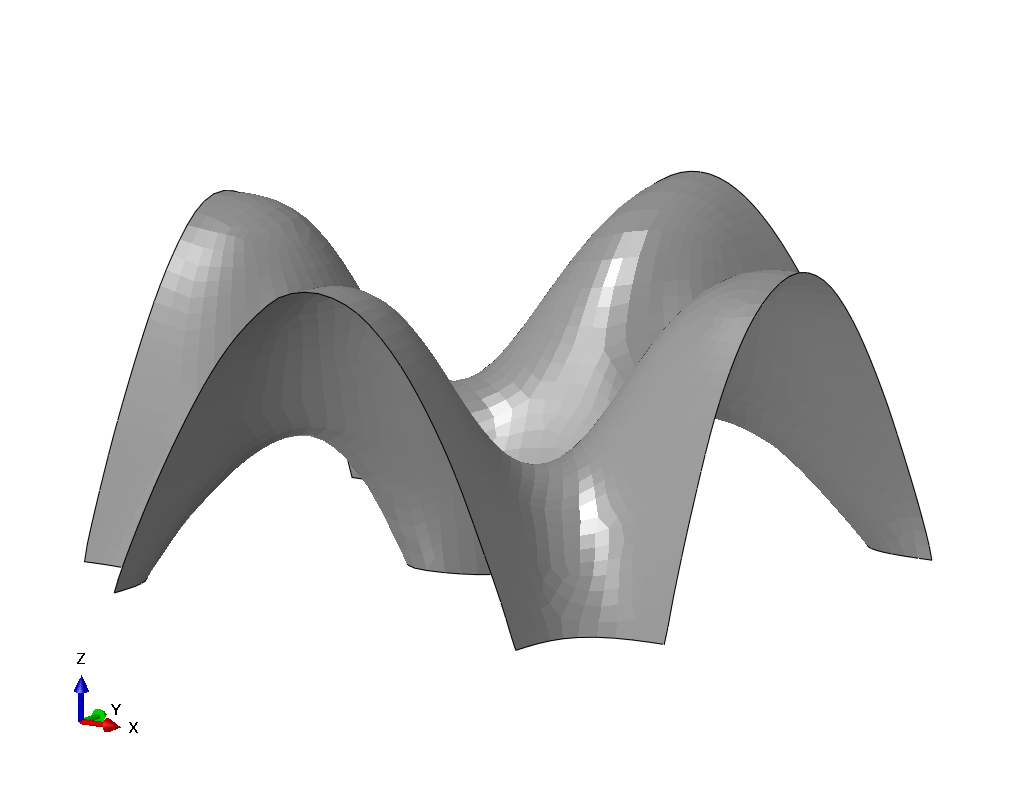}
    }\hspace{6mm}
    \subfloat[]{
        \includegraphics[width=0.22\textwidth]{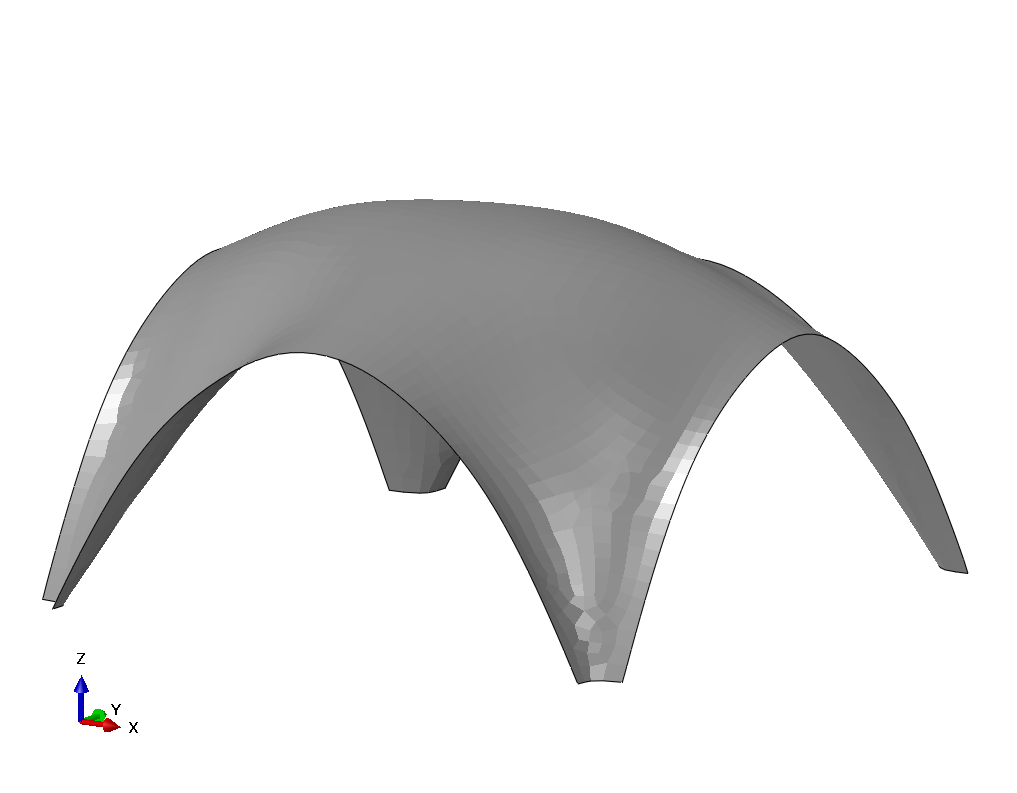}
    }
    \caption{Procedurally generated funicular shell geometries.}\label{fig:shells}
\end{figure}

A lattice modifier with a predefined number of control points was used to obtain smooth, funicular-type structures with varying heights and shapes, similar to the examples shown in Fig. \ref{fig:shells}. The last frame of the simulation was captured and the extremities of each structure were trimmed to avoid sharp corners and simulate physical supports.  The resulting mesh-based surface was exported as an STL file, subsequently converted into a parametric surface, and imported into Abaqus/Standard \cite{smith2009} for FEA.

\begin{figure}[!htbp]
    \centering
    \subfloat[]{
        \includegraphics[width=0.26\textwidth]{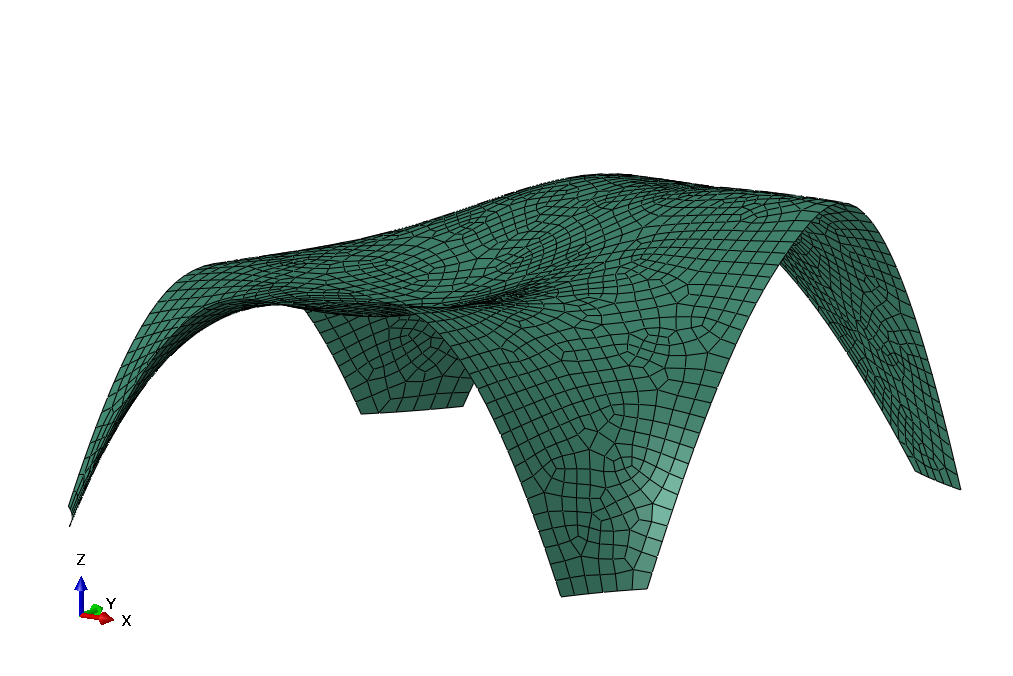}
    }%\hspace{4mm}
    \subfloat[]{
        \includegraphics[width=0.26\textwidth]{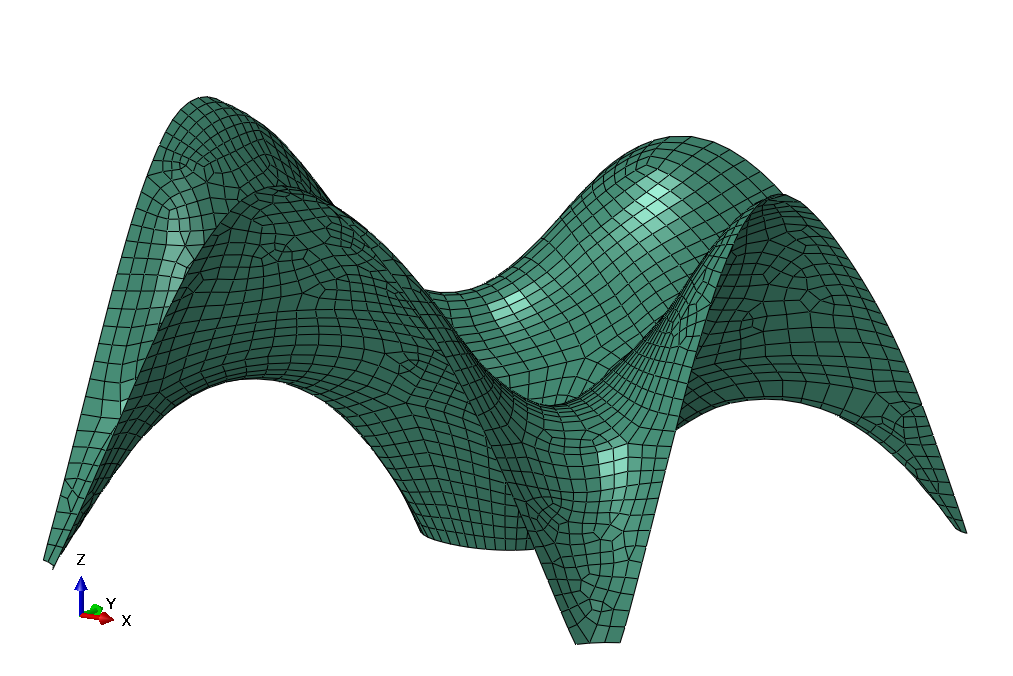}
    }
    \subfloat[]{
        \includegraphics[width=0.24\textwidth]{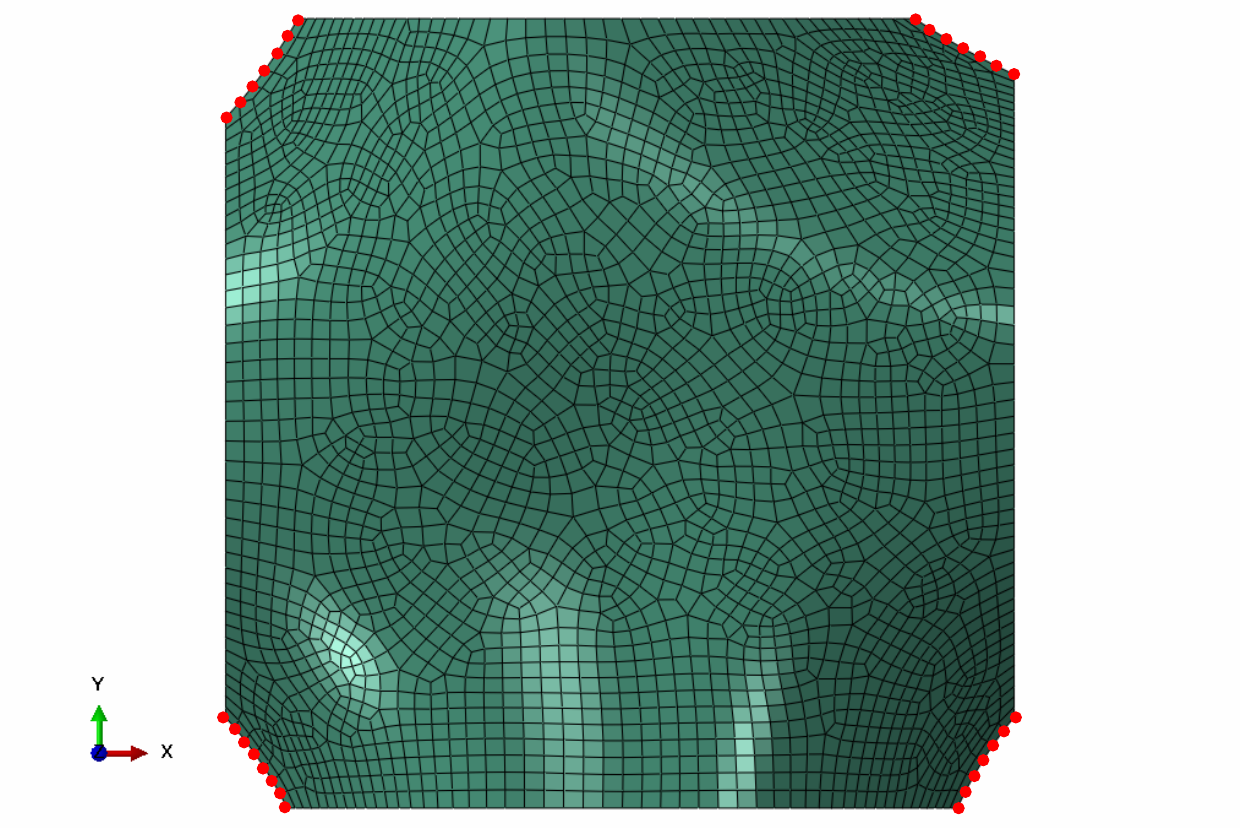}\label{fig_bc1}
    }%\hspace{-2mm}
    \subfloat[]{
        \includegraphics[width=0.24\textwidth]{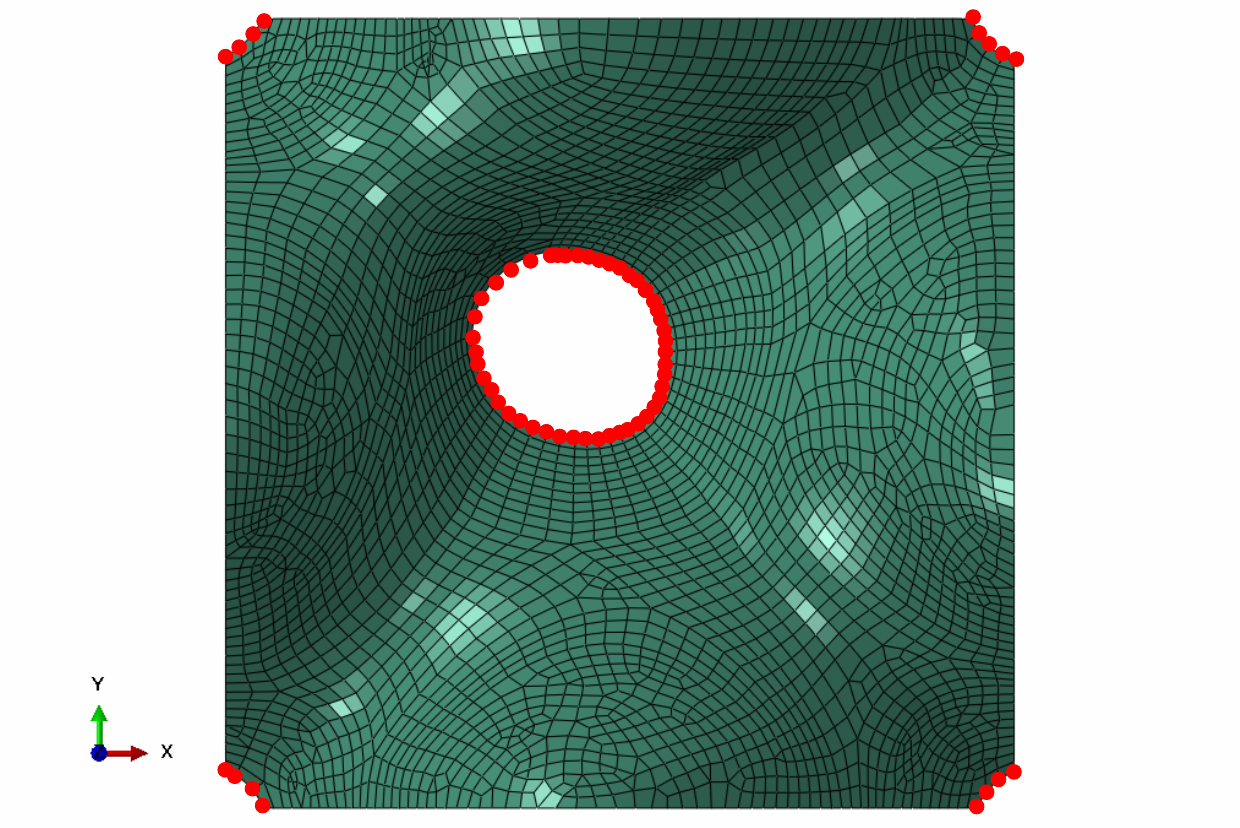}\label{fig_bc2}
    }
    \caption{Mesh densities obtained with the described mesh properties for surfaces (a) without central hole and (b) with a central hole. Points highlighted in red are nodes with fixed supports.}\label{fig:mesh}
\end{figure}

For the FEA, the shells were assumed to have thickness $h = 0.1~{m}$ and be made of a linear elastic material with Young modulus $E = 30~{GPa}$, a Poisson's ratio $\nu=0.2$ and density $\rho=2500~{kg/m^3}$. This hypothesis in no way invalidates the method for designing shells made from materials such as 3D-printed concrete. Like the more widely used methods for finding laminar shapes---such as the strut-and-tie method \cite{schlaich91}, the stress-fields method \cite{muttoni96} or the thrust network analysis \cite{block07}---this assumption of linearity is supported by the lower bound theorem in structural analysis. The geometries were discretized using a quad-dominated mesh composed of S4R and/or S3R elements, depending on the case. Multiple mesh convergence studies allowed the choice of an appropriate approximate global element size of 0.25 m, resulting in mesh densities comparable to those shown in Fig. \ref{fig:mesh}. Assuming isotropic material behaviour and isothermal conditions, a linear static analysis was performed under uniform self-weight loading, with gravitational acceleration $g = 9.81~{m/s^2}$, and fixed boundary conditions applied to the nodes at the extremities in contact with the ground (Figs. \ref{fig_bc1} and \ref{fig_bc2}). At the last time stage, the following variables were extracted: the nodal coordinates $x$, $y$ and $z$; the element area ${d}s$ and volume ${d}v$; the membrane factor $m_{{f}}$, the vertical load $f_z$, the membrane and flexural strain tensors, $\boldsymbol{\varepsilon}^m$ and $\boldsymbol{\varepsilon}^f$; the stress resultants $\boldsymbol{{F}}$ and bending moments $\boldsymbol{{M}}$; and finally, the membrane $\mathcal{W}_{\text{memb}}$, flexural $\mathcal{W}_{\text{flex}}$ and external work $\mathcal{W}_{\text{ext}}$.

%%%%%%%%%%%%%%%%%%%%%%%%%%%%%%%%%%%%
%% Geometry representation
%%%%%%%%%%%%%%%%%%%%%%%%%%%%%%%%%%%%
\subsection{Geometry representation}

\begin{figure}[!h]%
\centering
\includegraphics[width=0.4\textwidth]{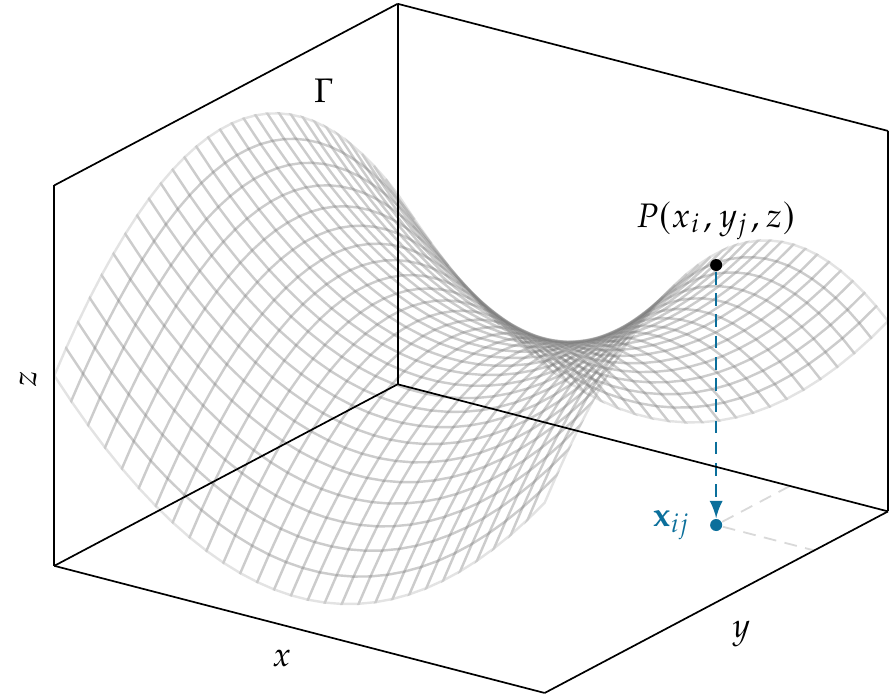}
\caption{The 3D surface $\Gamma$ is discretized into a 2D grid of pixels with coordinates $\boldsymbol{{x}}_{ij}$ and value $f(x_i , y_j , z)$, which is projected onto the 12 plane.}
\label{fig:projection}
\end{figure}

To feed the GAN with the generated data, an image-based representation of the shell structures was used. That is, the 3D shell structures were discretized and projected onto a 2D grid of pixels, where each pixel corresponds to the height of the shell at that point. Given the slenderness of these structures, their geometries can be approximated as a 2.5D curved surface $\Gamma$ in the domain $\Omega\subset\mathbb{R}^3$ and described by a level set function $f : \mathbb{R}^2 \rightarrow \mathbb{R}$, where $f(x, y) = z$ maps the height $z$ to a point $(x,y)$ in the shell geometry. The shells thus generated are smooth and continuous, such that each point $(x, y)$ corresponds to a unique height $z$, ensuring a one-to-one mapping from the 3D surface to its projection on the $xy$-plane (see Fig. \ref{fig:projection}). As such, it is possible to discretize the planar domain into a regular grid of $n_x \times n_y$ points. At each point of the grid, the height $z$ is saved essentially resulting in a image with the pixel values representing the elevation of the shell, or any other field of interest \cite{osher2004}.

For this work, the geometries were embedded in a square grid; a mask was used to define the holes. In particular, the variables were interpolated from the dense FEA mesh of points to a regular grid of $64\times64$ pixels. The full dataset was split into training (80\%), validation (10\%) and testing (10\%). The data was normalized using either Min-max or Z-score. Min-max normalization rescales the input data to a fixed range, typically $[-1,1]$, according to the formula:
\begin{equation}
    {{{x'}}} = 2\cfrac{{{x}}-x_{\text{min}}}{x_{\text{max}}-x_{\text{min}}}-1~,
\end{equation}
where ${{{x'}}}$ is the normalized values of the input feature ${{x}}$, $x_{\text{min}}$ and $x_{\text{max}}$ are the minimum and the maximum values of the input feature, respectively. On the other hand, the Z-score transforms the input data to have zero mean and unit variance, such that:
\begin{equation}
    {{{x'}}} = \cfrac{{{x}}-x_{\mu}}{x_s}~,
\end{equation}
where $x_{\mu}$ and $x_s$ are the mean and variance of the input feature ${{x}}$. For both techniques, the masked regions of the image data were not taken into account, as to not skew the statistics. The choice of one technique over the other was motivated by the results of a hyperparameter study (see Section \ref{ssc:hyper_study}).

%%%%%%%%%%%%%%%%%%%%%%%%%%%%%%%%%%%%
%% UNet
%%%%%%%%%%%%%%%%%%%%%%%%%%%%%%%%%%%%
\subsection{Physics-based Parallel UNet}
\label{ssc:unet}

\begin{figure}[h!]%
\centering
\includegraphics[width=0.4\textwidth]{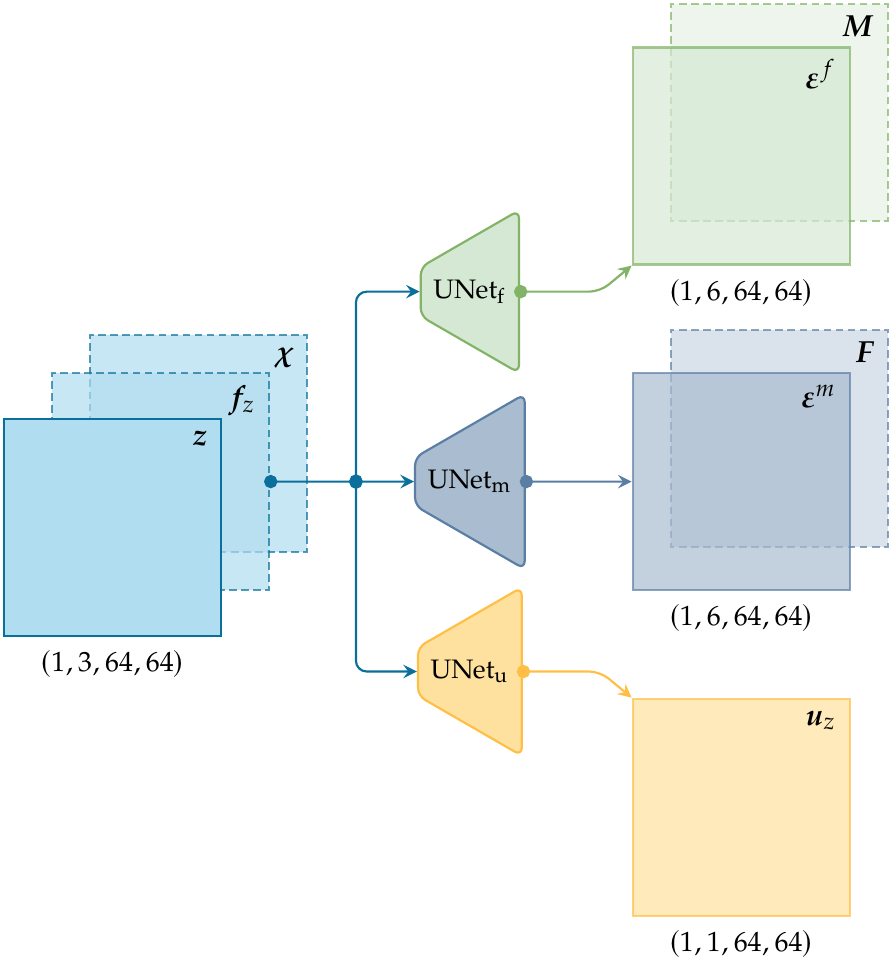}
\caption{PB-PUNet architecture to predict the behaviour of shell structures. }
\label{fig:unet}
\end{figure}

To learn the physical behaviour of the shell structures, an implementation based of the UNet architecture described by Ronneberger \textit{et al.} \cite{ronneberger2015} was adapted here to a physics-informed regression task. The base architecture is essentially a U-shaped symmetric encoder-decoder with skip connections, where the contracting path employs convolutional blocks with ReLU activation functions and batch normalization, followed by pooling layers to downsample the feature maps. The expansive path mirrors this process, using transposed convolutions for upsampling and merging feature maps from the encoder to refine segmentation boundaries. Multiple parallel UNet subnets are dedicated to predicting each part of the solution fields as depicted in Fig. \ref{fig:unet}, hence the designation Physics-based Parallel UNet (PB-PUNet). The approach effectively decouples fields governed by different physical laws or distinct orders of magnitude, to improve the model performance \cite{sun2020,guo2020}. Motivated by a careful optimization of the model's architecture, three identical parallel subnets were employed

The model takes as input a three-channel image $\boldsymbol{{x}}$, expressed as:
\begin{equation}
    \boldsymbol{{x}}=\bigl[z, f_z, \chi \bigr],
\end{equation}
where $z$ is the height of the shell and $f_z$ is the uniform vertical self-weight loading. Given that the image-based representations of the shells may contain empty pixel regions, the input is augmented by a binary mask $\chi$ defined, such that:
\begin{equation}
    \chi = \left\{(i,j)\vert{} z_{i,j}>0\right\},
\end{equation}
with $(i,j)$ a pair of pixel coordinates in image channel $z$, and assuming values of 1 for the valid pixels ($z>0$) and 0 elsewhere. The binary mask provides a strong inductive bias to prevent the convolution kernels from treating zero-valued placeholders as meaningful signal, thus avoiding these empty regions from bleeding into the features of the actual image. Each subnet predicts the pixel-wise vertical displacement ${{u}}_z$, membrane strains $\boldsymbol{\varepsilon}^m$ and stress resultants $\boldsymbol{{F}}$, flexural strains $\boldsymbol{\varepsilon}^f$ and bending moments $\boldsymbol{{M}}$, respectively. These solution fields are concatenated into a thirteen-channel image $\boldsymbol{{y}}$, such that:
\begin{equation}
    \boldsymbol{{y}}=\left[u_z, \varepsilon_{11}^{m}, \varepsilon_{22}^{m}, \varepsilon_{12}^{m}, F_{11}, F_{22}, F_{12}, \varepsilon_{11}^{f}, \varepsilon_{22}^{f}, \varepsilon_{12}^{f}, M_{11}, M_{22}, M_{12}\right],
\end{equation}
which are necessary to evaluate the membrane factor distribution ${{m}}_{{f}}$.

%%%%%%%%%%%%%%%%%%%%%%%%%%%%%%%%%%%%
%% AD-DCGAN architecture
%%%%%%%%%%%%%%%%%%%%%%%%%%%%%%%%%%%%
\subsection{AD-DCGAN arquitecture}

\begin{figure}[h!]%
\centering
\includegraphics[width=0.75\textwidth]{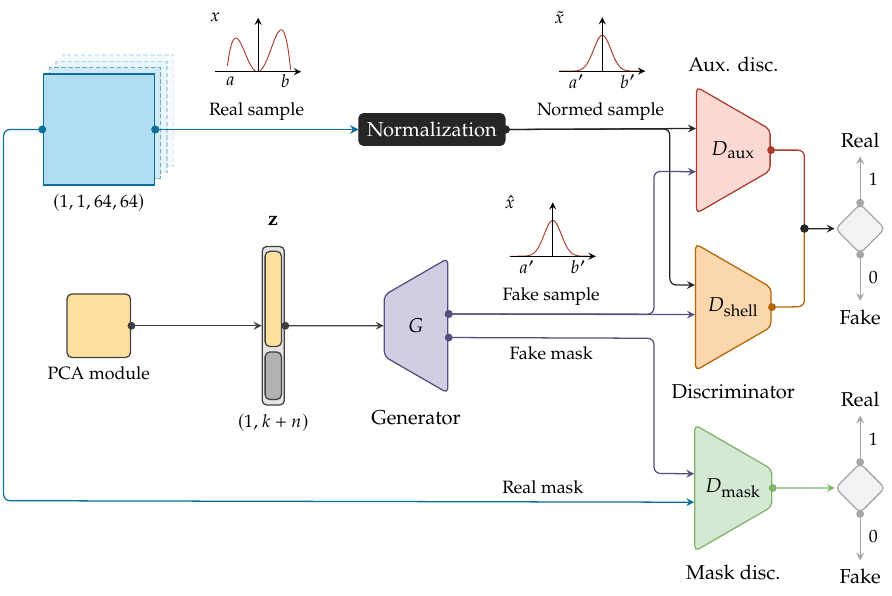}
\caption{AD-DCGAN architecture. $G$ converts a PCA-guided latent vector $\boldsymbol{{z}}$ into synthetic images and binary masks. Both $D_{\text{shell}}$ and $D_{\text{mask}}$ classify the generated images and binary masks, respectively, as real or fake.  $D_{\text{aux}}$ is a pre-trained PINN that evaluates the physical compliance of the generated shells through the membrane factor distribution ${{m}}_{{f}}$.}
\label{fig:ad_dcgan}
\end{figure}

A custom architecture based on the original deep convolutional generative adversarial network, DCGAN, by Radford \textit{et al.} \cite{radford2016} is hereby proposed for the generation of funicular shells, as depicted in Fig. \ref{fig:ad_dcgan}. The model is designed to operate with image-based representations of the shell geometries and employs an auxiliary discriminator ($D_{\text{aux}}$), hence the designation AD-DCGAN. The former is a pre-trained PB-PUNet model that evaluates the physical compliance of the newly generated shells according to the corresponding membrane factor distribution ${{m}}_{{f}}$. Based on the work of Gao \textit{et al.} \cite{gao2025}, the AD-DCGAN also incorporates a Principal Component Analysis (PCA) preprocessing module, to guide the generator ($G$) input based on the top $k$ principal components extracted from the real data, instead of relying purely on random noise. The approach mitigates mode collapse caused by random noise inputs and reduces the computational complexity of the generator. As a result, a low-dimensional feature space of enhanced latent variables $\boldsymbol{{z}}_{k+n}$ is constructed based on the $k$ extracted features and augmented by $n$ random noise components taken from a normal distribution with zero mean and unit variance. The AD-DCGAN is trained in a minmax game, where the $G$ aims to produce realistic shell structures that fool the discriminator ($D_{\text{shell}}$), while the latter aims to distinguish between real and generated shells.

\begin{figure}[!htbp]%
\centering
\includegraphics[width=0.95\textwidth]{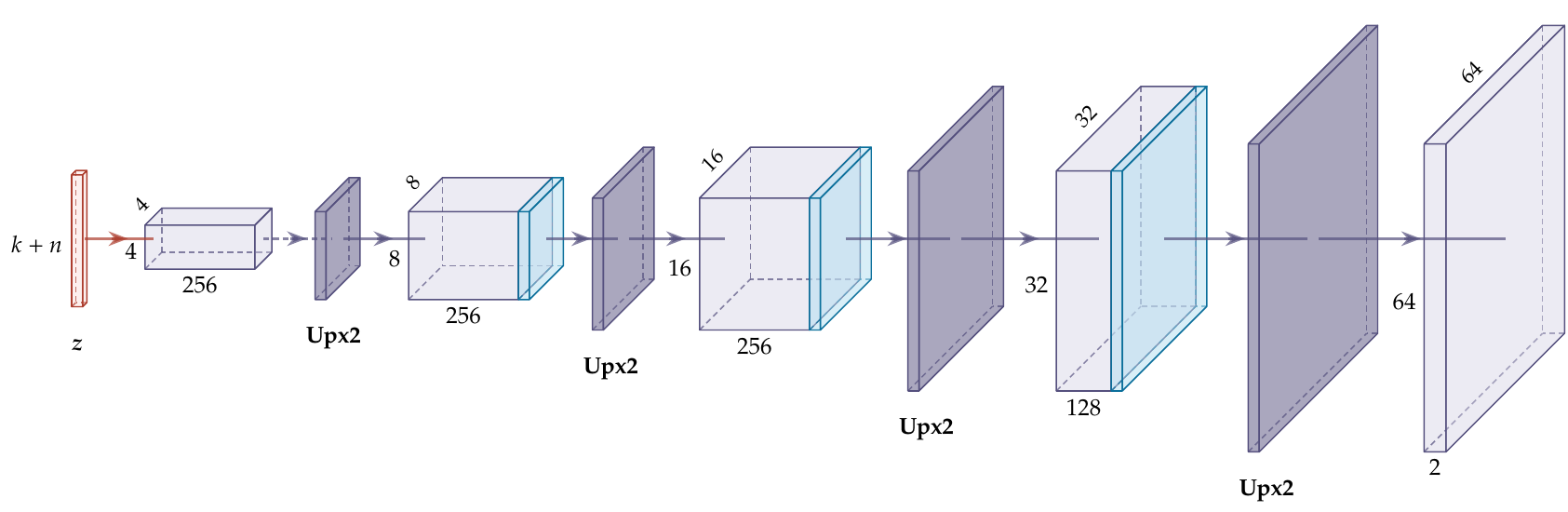}
\caption{AD-DCGAN generator architecture for shell design. A $(k+n)$-dimensional PCA-guided latent vector $\boldsymbol{{z}}$ is projected to a small feature map, which is converted to a $2\times64\times64$ pixel image by a series of four convolutions, each preceded by an upsample block. The light blue blocks, represent batch normalization and Leaky ReLU.}
\label{fig:gen}
\end{figure}

Following the detailed representation from Fig. \ref{fig:gen}, the generator is essentially a 2D-CNN that generates pixel-wise representations of the shells taking as input the PCA-guided latent vector $\boldsymbol{{z}}$ and projecting it into a $256 \times 4 \times 4$ feature map with a fully connected layer. Subsequently, four upsampling steps and convolution operations are applied, sequentially doubling the spatial dimensions to obtain an intermediate array of shape $128\times32\times32$. Batch normalization is introduced after each convolution to reduce the impact of internal covariate shift and help with gradient flow \cite{radford2016,ioffe2015}. Then, the Leaky ReLU activation is used with a small negative slope of 0.2. Finally, an upsample step followed by a convolution operation double the spatial dimensions of the previous output to obtain a final array of $2\times64\times64$. The final image is obtained by passing one of the channels through a hyperbolic tangent (Tanh) activation function, while a binary mask is obtained by passing the remaining channel through a sigmoid activation. The mask depicts the valid and invalid regions for the generated shell geometries. As such, a mask discriminator ($D_{\text{mask}}$) was also used to evaluate the generated masks, thus guiding the generator to correctly depict such regions. This was necessary given the way the forward pass of the PB-PUNet model was structured (see Section \ref{ssc:unet}). 
 
\begin{figure}[h!]%
\centering
\includegraphics[width=0.6\textwidth]{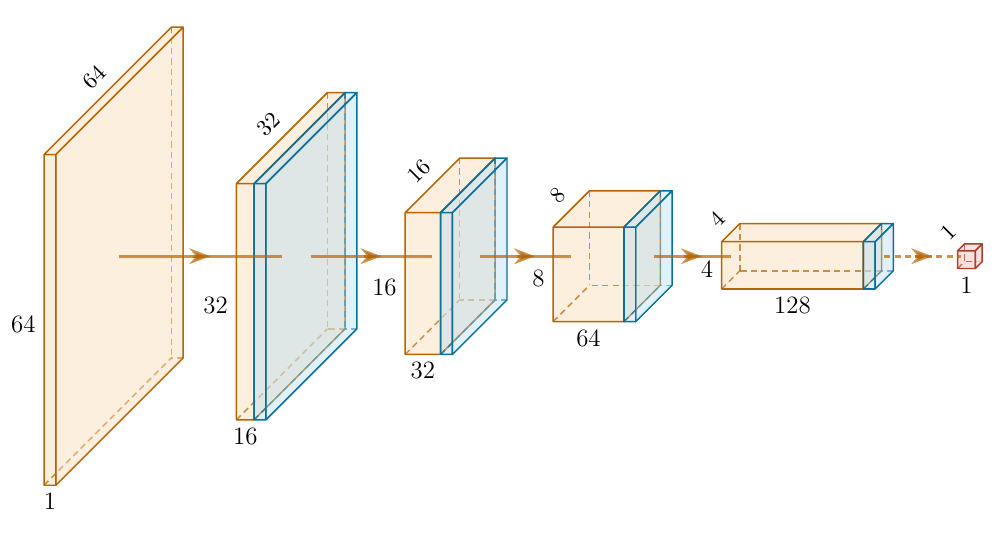}
\caption{AD-DCGAN discriminator architecture for shell design. A series of four convolutions progressively downsample the input image and the result is passed through a final fully-connected layer to produce a single probability. The light blue blocks represent Leaky ReLU, dropout and batch normalization.}
\label{fig:disc}
\end{figure}

The discriminators $D_{\text{shell}}$ and $D_{\text{mask}}$ share the detailed representation from Fig. \ref{fig:disc}. Here, the input is a single $64\times64$ image which is progressively downsampled by a series of four convolutions using a $3\times3$ kernel and a stride of 2, to obtain an intermediate output of shape $128\times4\times4$. After each convolution layer, LeakyReLU activation, dropout and batch normalization are introduced. Finally, the previous output is mapped to a single scalar through a fully-connected layer. Spectral normalization is employed on the convolution and fully-connected layers.

%%%%%%%%%%%%%%%%%%%%%%%%%%%%%%%%%%%%
%% Training
%%%%%%%%%%%%%%%%%%%%%%%%%%%%%%%%%%%%
\subsection{Training procedure}
\subsubsection{PB-PUNet}

The PB-PUNet model was trained using the Pytorch library following a standard supervised learning approach with
labelled data. Training was carried out for all subnets simultaneously, with the Adam optimizer using a learning rate of $1\times10^{-4}$. An early-stopping criterion was implemented to prevent overfitting and interrupt training if the validation loss did not improve after 150 epochs. The model was trained to minimize the loss function:
\begin{equation}
    \mathcal{L}_{\text{PB-PUNet}} = \left\Vert\boldsymbol{{y}}-\hat{\boldsymbol{{y}}}\right\Vert^2_2 + \lambda_{\text{phys}}\cdot\mathcal{\delta P}(\hat{\boldsymbol{{y}}}),
\end{equation}
where $\boldsymbol{{y}}$ are the ground-truth values, $\hat{\boldsymbol{{y}}}$ the predicted values, $\lambda_{\text{phys}}$ a regularization factor and $\mathcal{\delta P}$ the physics loss, based on the balance described by Eq. (\ref{eq:balance}). A baseline model was initially developed using the hyperameters listed in Table \ref{tab:baseline}.

\begin{table}[!ht]
\centering
\caption{Hyperparameters for the PB-PUNet baseline model.\label{tab:baseline}}
\begin{tabular}{@{}lc@{}}
\toprule
\multicolumn{2}{c}{\textbf{Baseline}} \\ \midrule
Batch size & 8 \\
Learning rate & $1\times10^{-4}$ \\
Reg. factor ($\lambda_{\text{phys}}$) & 1.0 \\
Weight initialization & Default \\
Data normalization & MinMax \\
Batchnorm & yes \\ 
Dropout & no \\\bottomrule
\end{tabular}
\end{table}

\subsubsection{AD-DCGAN}

The AD-DCGAN model was trained with two Adam optimizers, both defined with $\beta_1$ and $\beta_2$ parameters of $0.5$ and $0.999$, and learning rates of $1.0\times{}10^{-4}$ and $2.0\times{}10^{-4}$ for both discriminators and the generator, respectively. The batch size was set to $8$ and the model was trained for $2000$ epochs. The total generator loss was defined as:
\begin{equation}
\mathcal{L}_{{G_{\text{total}}}} = \mathcal{L}_{{{G_{\text{shell}}}}}^{\text{fake}} + \mathcal{L}_{{{G_{\text{mask}}}}}^{\text{fake}} + \mathcal{L}_{\text{FM}}+\lambda_{{D_{\text{aux}}}}\cdot\mathcal{L}_{D_{\text{aux}}}^{\text{fake}}~,
\end{equation}
where $\mathcal{L}_{\text{FM}}$ is the feature matching loss term defined in Section \ref{ssc:feat_match} and $\lambda_{{D_{\text{aux}}}}$ is a regularization factor for the auxiliary discriminator term. The remaining terms are defined such that:
\begin{align}
\mathcal{L}_{{{G_{\text{shell}}}}}^{\text{fake}}&=\mathcal{E}\bigl({D_{\text{shell}}}({{G_{\text{shell}}}}(\boldsymbol{{z}})), 1\bigr),\nonumber\\[2.25mm]
\mathcal{L}_{{{G_{\text{mask}}}}}^{\text{fake}}&= \mathcal{E}\bigl({{D_{\text{mask}}}}({{G_{\text{mask}}}}(\boldsymbol{{z}})), 1\bigr),\\[2.25mm]
\mathcal{L}_{{D_{\text{aux}}}}^{\text{fake}}&=\mathcal{E}\bigl({{D_{\text{aux}}}}({G_{\text{shell}}}(\boldsymbol{{z}})),1\bigr),\nonumber
\end{align}
with $\mathcal{E}$ an error function, herein taken as the binary cross-entropy loss (BCE). The total discriminator loss was defined as:
\begin{equation}
    \mathcal{L}_{{D_{\text{total}}}} = \mathcal{L}_{{D_{\text{shell}}}} + \mathcal{L}_{{D_{\text{mask}}}} + \mathcal{L}_{{D_{\text{aux}}}}~,
\end{equation}
with each term assuming the general form:
\begin{equation}
    \mathcal{L}_{{D}} = \frac{1}{2}\Bigl( \mathcal{L}_{{D}}^{\text{real}}+\mathcal{L}_{{D}}^{\text{fake}} \Bigr)~.
\end{equation}
According to their corresponding discriminator, $\mathcal{L}_{{D}}^{\text{real}}$ and $\mathcal{L}_{{D}}^{\text{fake}}$ are defined as follows:
\begin{equation}
\begin{aligned}
    \mathcal{L}_{{D_{\text{shell}}}}^{\text{real}} &= \mathcal{E}\bigl({D_{\text{shell}}}(\mathbf{x}), X\bigr), &
    \mathcal{L}_{{D_{\text{shell}}}}^{\text{fake}} &= \mathcal{E}\bigl({D_{\text{shell}}}({G_{\text{shell}}}(\mathbf{z})), 0\bigr),\\[2.25mm]
    \mathcal{L}_{{D_{\text{mask}}}}^{\text{real}} &= \mathcal{E}\bigl({D_{\text{mask}}}(\boldsymbol{\chi}), X\bigr), &
    \mathcal{L}_{{D_{\text{mask}}}}^{\text{fake}} &= \mathcal{E}\bigl({D_{\text{mask}}}({G_{\text{mask}}}(\mathbf{z})), 0\bigr)~,\\[2.22mm]
    \mathcal{L}_{{D_{\text{aux}}}}^{\text{real}} &= \mathcal{E}\bigl({D_{\text{aux}}}(\mathbf{x}), X\bigr), &
\mathcal{L}_{{D_{\text{aux}}}}^{\text{fake}} &= \mathcal{E}\bigl({D_{\text{aux}}}({G_{\text{shell}}}(\mathbf{z})), 0\bigr),
\end{aligned}
\end{equation}
where $\boldsymbol{{x}}$ and $\boldsymbol{\chi}$ are ground truth samples and binary masks, respectively. Following the indications from Salimans \textit{et al.} \cite{salimans2016}, label smoothing is applied, thus the real labels are replaced with a random number $X$, such that $X\sim\mathcal{U}(0.8,1.0)$. This prevents the discriminator from assigning near‑zero loss to real data, which can lead to vanishing gradients and stall the generator learning. Assigning a smooth value to the real label, allows the discriminator to provide more informative gradients, encouraging the generator to produce samples that more closely resemble the true data distribution.

%%%%%%%%%%%%%%%%%%%%%%%%%%%%%%%%%%%%
%% Validation
%%%%%%%%%%%%%%%%%%%%%%%%%%%%%%%%%%%%
\subsection{Validation procedure}
\subsubsection{PB-PUNet}

The predictive performance of the PB-PUNet model is evaluated using the Relative Root Mean Square Error (RRMSE) metric, which is a dimensionless form of the Root Mean Square Error (RMSE) normalized by the $q$-norm of the ground truth values, such that \cite{despotovic2016}:
\begin{equation}
    RRMSE_q = \sqrt{\cfrac{1}{N}\left(\cfrac{\left\Vert\boldsymbol{{y}}-\hat{\boldsymbol{{y}}}\right\Vert^2_2}{\left\Vert\boldsymbol{{y}}\right\Vert_q}\right)}.
\end{equation}
The $q$-norm is taken either as the maximum of the infinity norm ($q=\infty$) or the Euclidean
norm ($q$ = 2). If the outliers are moderate compared to the average value of the ground truth, the infinity norm provides a conservative estimate to evaluate the model accuracy, normalizing the errors by the maximum local value of the ground truth. However, if the outliers are more than a few orders of magnitude distant from the average value of the ground truth, the Euclidean norm provides a more balanced estimate, normalizing the errors by the square of the average value of the ground truth.

To evaluate the PB-PUNet model, feature sets are defined grouping the solution fields into different categories, such as: displacements (${{u}}_z$), membrane strains ($\boldsymbol{\varepsilon}^m$), stress resultants ($\boldsymbol{{F}}$), flexural strains ($\boldsymbol{\varepsilon}^f$) and bending moments ($\boldsymbol{{M}}$). An additional feature set (total) is also added to provide a single metric as a measure of the predictive performance that groups all the solution fields.

The normalized error can be computed over the pixels of every sample in a given data split (i.e. training, validation or testing) and over every feature in the feature sets. The normalized errors are squared and averaged over every sample in the
in the validation and testing data to obtain scalar feature-wise RRMSE values. The dimensionless nature of the RRMSE allows for a direct comparison of the predictive performance across different
feature sets and models. To gain insights into the spatial distribution of the errors in the
predicted solution fields, we also evaluated a pixel-wise RRMSE computed per sample and visualized as a heatmap image of the error over the shell domain.

\subsubsection{AD-DCGAN}

The metrics used to evaluate regressive models are not applicable to GANs. Instead, these models are typically assessed using metrics that rely on third-party classifiers pre-trained on benchmark datasets. The Fréchet Inception Distance (FID) \cite{heusel2018} is a widely used metric that compares the distribution of generated samples to that of real data by extracting features from the penultimate layer of the Inception v3 model \cite{szegedy2015} pre-trained on ImageNet dataset \cite{deng2009}.

In this study, however, the AD-DCGAN generates image-based representations of physical fields that do not share any natural features with the ImageNet dataset. As a result, FID was deemed unsuitable for evaluating model performance. Additionally, training Inception v3 would require substantial computational resources and data, which was not feasible in the context of this project. To address this, we propose leveraging the learned feature representations of the AD-DCGAN discriminator. The latter is essentially a binary classifier trained to distinguish real from generated samples. By extracting and flattening the features from its penultimate layer, a high-dimensional latent representation of each input image can be obtained, analogous to the FID.

To visualize how the generated samples cover the distribution of real samples in the latent manifold, several dimensionality reduction techniques can be considered. t-SNE \cite{vandermaaten2008} is a nonlinear method that preserves local structure but is stochastic, making it less suitable for reproducible analyses. UMAP \cite{mcinnes2020} is faster, preserves both local and global structure, and is non-stochastic; however, its output is highly sensitive to a wide range of hyperparameters, making it harder to fine tune. Kernel PCA \cite{scholkopfl1998}, the nonlinear extension of principal component analysis, effectively captures the nonlinear spatial structure. It was ultimately chosen to evaluate the AD-DCGAN latent manifold because it is deterministic, computationally efficient and depends on a very limited number of hyperparameters.

Additionally, a set of generated surfaces are imported into Abaqus/Standard software. The FEA is performed under the same conditions used to analyse the training dataset (see Section \ref{ssc:database}). The final membrane factor distribution is then evaluated against the one provided by the PB-PUNet for that set of surfaces.

%%%%%%%%%%%%%%%%%%%%%%%%%%%%%%%%%%%%
%% Results
%%%%%%%%%%%%%%%%%%%%%%%%%%%%%%%%%%%%
\section{Results}\label{sc:results}
%%%%%%%%%%%%%%%%%%%%%%%%%%%%%%%%%%%%
%% Results - PINN
%%%%%%%%%%%%%%%%%%%%%%%%%%%%%%%%%%%%
\subsection{PB-PUNet}

\subsubsection{Hyperparameter tuning}\label{ssc:hyper_study}

Hyperparameter tuning was performed to find the optimal combination of parameters for the PB-UNet taking into account: the type of data normalization, the weight initialization approach, the use of Batchnorm or Groupnorm, the effect of dropout and the magnitude of the physics regularization factor $\lambda_{\text{phys}}$. The study compared the RRMSE$_{\infty}$ of the predicted state values from five different models against the baseline, and the corresponding results are shown in Table \ref{tab:results}. In order to streamline the process, the hyperparameter tuning was performed on a smaller set of data with did not contain any surfaces with a central hole. Nonetheless, the final model was trained on the complete dataset using the optimized parameters found here, as those conclusions were still considered valid. 

\begin{table}[!h]
\centering
\caption{Overview of the different models under study: main specifications and RRMSE$_{\infty}$ for each predicted output. The percent values in between parenthesis are the relative errors based on the baseline model.}
\resizebox{\textwidth}{!}{%
\begin{tabular}{ccccccc}
\toprule
\multicolumn{1}{l}{\multirow{2}{*}{}} & \multicolumn{6}{c}{\textbf{Models}} \\ \midrule
\multicolumn{1}{l}{} & \textbf{Baseline} & \textbf{Model A} & \textbf{Model B} & \textbf{Model C} & \textbf{Model D} & \textbf{Model E} \\ \midrule
Data norm. & Min-max & Z-score & Z-score & Z-score & Z-score & Z-score \\
Weight init. & default & default & Kaiming & Kaiming & Kaiming & Kaiming \\
Batchnorm & yes & yes & yes & no & no & no \\
Groupnorm & no & no & no & yes & yes & yes \\
Dropout & no & no & no & no & 0.3 / 0.2 & 0.2 / 0.1 \\
$\lambda_{\text{phys}}$ & 1 & 1 & 1 & 1 & 1 & 0.1 \\ \midrule 
& \multicolumn{6}{c}{\textbf{RRMSE}\textsubscript{$\infty$}} \\ \midrule
${{m}}_{{f}}$ & 62.741 & 0.046 (-99.93\%) & 0.048 (-99.93\%) & 0.044 (-99.93\%) & 0.054 (-99.91\%) & 0.048 (-99.92\%) \\
${{u}}_z$ & 0.134 & 0.109 (-18.65\%) & 0.084 (-37.31\%) & 0.074 (-44.77\%) & 0.077 (-42.54\%) & 0.050 (-62.69\%) \\
$\boldsymbol{\varepsilon}^m$ & 0.103 & 0.020 (-80.58\%) & 0.019 (-81.55\%) & 0.017 (-83.49\%) & 0.028 (-72.82\%) & 0.022 (-78.64\%) \\
$\boldsymbol{{F}}$ & 0.094 & 0.016 (-82.97\%) & 0.015 (-84,04\%) & 0.014 (-85.10\%) & 0.023 (-75.53\%) & 0.017 (-81.91\%) \\
$\boldsymbol{\varepsilon}^f$ & 0.066 & 0.057 (-13.63\%) & 0.053 (-19.70\%) & 0.058 (-12.12\%) & 0.086 (+30.30\%) & 0.054 (-18.18\%) \\
$\boldsymbol{{M}}$ & 0.065 & 0.056 (-13.85\%) & 0.053 (-18.46\%) & 0.056 (-13.85\%) & 0.083 (+27.69\%) & 0.053 (-18.46\%) \\
\textbf{Total} & 0.094 & 0.016 (-82.98\%) & 0.015 (-84.04\%) & 0.014 (-85.11\%) & 0.023 (-75.53\%) & 0.017 (-81.91\%) \\ \bottomrule
\end{tabular}%
}
\label{tab:results}
\end{table}

Observing the results, the baseline model with Min-max normalization struggles to provide an accurate prediction of the membrane factor distribution, showing abnormally high error compared to the other variables. The issue was solved using Z-score normalization (Model A) resulting in a $99.93\%$ error reduction for the membrane factor. Overall, in comparison to the baseline, Model A also offers better predictive accuracy across the remaining fields, with special emphasis on the membrane strains $\boldsymbol{\varepsilon}^{m}$ and the internal forces $\boldsymbol{{F}}$ showing an approximate $80\%$ improvement over the baseline. 

According to Ronneberger \textit{et al}. \cite{ronneberger2015}, ideally the weights should be initialized such that each feature map in the network has approximately unit variance. To achieve this, the authors suggest an initialization technique equivalent to Kaiming normal initialization applied to the convolutional layers and setting the initial bias terms to 0. Employing these modifications (Model B) the RRMSE$_{\infty}$ values show incremental improvements in predictive performance compared to Model A, with the exception being the displacement field ${{u}}_z$ which experiences a significant $37.31\%$ improvement compared to the baseline. 

The baseline model employed Batchnorm to mitigate the internal covariate shift, however, this technique is not suitable for batch sizes lower than 16, as in such cases the batch statistics are not correctly estimated. As a result, Groupnorm was employed instead (Model C), as it is advisable for smaller batches. In general, the RRMSE$_{\infty}$ values show incremental improvements compared with Model B, with the exception again being the displacement field ${{u}}_z$ with a $44.77\%$ error reduction.

The PB-PUNet naturally exhibits a severe tendency to overfit the data. Despite that fact, the baseline model did not contemplate any form of regularization in order to mitigate it. As such, dropout layers were introduced to both the encoder and decoder components of the network. For the decoder part, the dropout rate should always be lower than the one applied to the encoder. As an exploratory exercise, dropout rates of 0.3 and 0.2 (Model D), and 0.2 and 0.1 (Model E) were applied to the encoder and decoder, respectively. Consequently, some RRMSE$_{\infty}$ values actually increased for Model D, with the highest being the flexural strains $\boldsymbol{\varepsilon}^f$ and the bending moments $\boldsymbol{{M}}$. On the other hand, Model E achieves similar results, however, reduced dropout rates allowed to mitigate the higher RRMSE$_{\infty}$ values previously seen with Model D. Further improvements were obtained by setting $\lambda_{\text{phys}}$ equal to 0.1, thus allowing Model E to achieve the highest predictive accuracy.

\subsubsection{Training dynamics and feature-wise errors}

\begin{figure}[!htbp]
    \centering
    \subfloat[]{
        \includegraphics[width=0.5\textwidth]{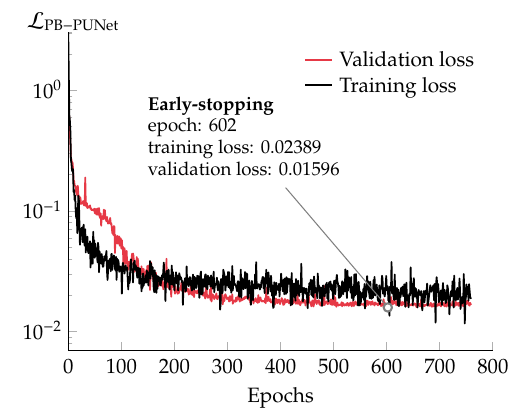}\label{fig:loss_train}
    }%\hspace{6mm}
    \subfloat[]{
        \includegraphics[width=0.5\textwidth]{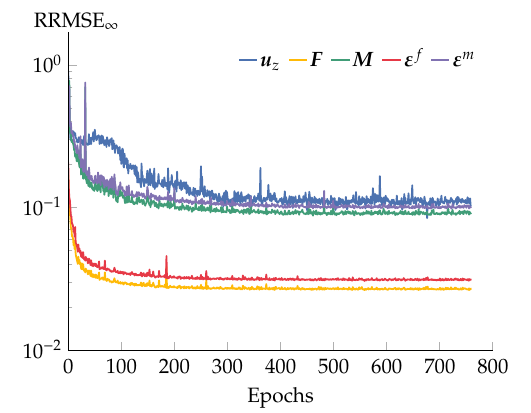}\label{fig:rrmse_train}
    }\hspace{6mm}
    \caption{Learning curves for the optimized PB-PUNet model trained on the full dataset: (a) training and validation losses, and (b) feature-wise average $\text{RRMSE}_{\infty}$ errors accumulated over the batches, evaluated on the validation split.}\label{fig:learn_curves}
\end{figure}

For the final PB-PUNet model trained with the optimized parameters on the complete dataset, both training and validation losses sharply decrease for the first 50 epochs, followed by a gradual evolution for the remaining epochs of training (Fig. \ref{fig:loss_train}). The training dynamic shows a slight tendency for the model to underfit the data, with the validation curve appearing slightly under the training curve. This indicates the model could possibly benefit from further adjustments in the dropout rates to invert this tendency; nevertheless, there is no substantial gap between both curves for the phenomenon to be considered problematic. The early-stopping was triggered after 602 epochs, saving the model state at this point with a train and validation loss of 0.02389 and 0.01596, respectively. 

The feature-wise RRMSE$_{\infty}$ curves evaluated in the validation set (see Fig. \ref{fig:rrmse_train}) show a similar dynamic, sharply decreasing at the beginning of training and experiencing a gradual evolution afterwards. In contrast, for the vertical displacement ${{u}}_z$ the error curve evolves more gradually with a noisy pattern, indicating a certain difficulty of the model to learn this state variable. The feature-wise $\text{RRMSE}_{\infty}$ errors obtained over the test set (Table \ref{tab:rrmse_test}) further confirm this tendency with the vertical displacement ${{u}}_z$ registering the highest error (10.244\%). All of the remaining feature sets register errors lower than 10\%, with the highest being the membrane factor ${{m}}_{{f}}$ (5.566\%), the bending moments $\boldsymbol{{M}}$ (6.403\%) and the flexural strains $\boldsymbol{\varepsilon}^f$ (7.205\%). Nevertheless, these errors account to a total RRMSE$_{\infty}$ of 2.198\%.

\begin{table}[!htbp]
\centering
\caption{Feature-wise RRMSE errors of the optimized PB-PUNet model evaluated on the test set.}
\begin{tabular}{ccc}
\toprule
\textbf{Feature set} & $\text{{RRMSE}}_2$ (\%) & $\text{{RRMSE}}_{\infty}$ (\%) \\ \midrule
${{m}}_{{f}}$ & 0.129 & 5.566\\
${{u}}_z$ & 0.396 & 10.344\\
$\boldsymbol{\varepsilon}^m$ & 0.148 & 2.511\\
$\boldsymbol{{F}}$ & 0.147 & 2.197\\
$\boldsymbol{\varepsilon}^f$ & 0.362 & 7.205\\
$\boldsymbol{{M}}$ & 0.37 & 6.403\\\midrule
\textbf{Total} & 0.147 & 2.198\\\bottomrule
\end{tabular}
\label{tab:rrmse_test}
\end{table}

\subsubsection{Inference on test data}

\begin{figure}[!htbp]%
\centering
\includegraphics[width=\textwidth]{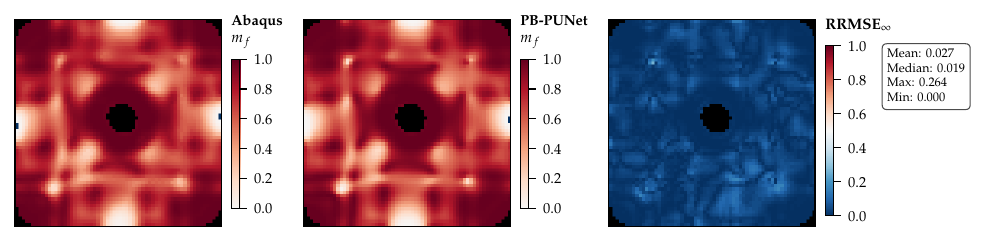}
{\caption{Pixel-wise RRMSE$_{\infty}$ prediction error of the PB-PUNet on the membrane factor distribution ${{m}}_{{f}}$ for the first sample of the test set.}
\label{fig:mf}}
\end{figure}

\begin{figure}[!htbp]%
\centering
\includegraphics[width=\textwidth]{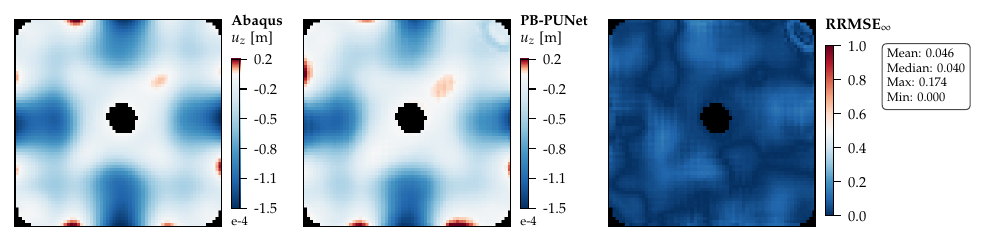}
{\caption{Pixel-wise RRMSE$_{\infty}$ prediction error of the PB-PUNet on the vertical displacement ${{u}}_z$ for the first sample of the test set.}
\label{fig:uz}}
\end{figure}

In this section, the contour maps of the state variables obtained with the PB-PUNet model, for the first sample of the testing set are analysed and compared to the FEA solution. Given the high amount of output variables, for the sake of clarity and readability, the authors choose to display here the results obtained for the membrane factor distribution ${{m}}_{{f}}$ (Fig. \ref{fig:mf}) and the vertical displacement ${{u}}_z$ (Fig. \ref{fig:uz}), keeping in mind that the former is the variable of interest for this project and the latter was the one registering the highest error magnitude over the test set. The remaining state variables are relegated to \ref{ap:pinn}, nevertheless, a general overview of the results obtained for all of them is provided. 

Analysing the results for the membrane factor ${{m}}_{{f}}$, the PB-PUNet model approximates the global FEA solution quite well, exhibiting overall low RRMSE$_{\infty}$ across the shell domain, achieving mean and median errors as low as approximately 2\%. The error map shows the higher magnitude errors tend to be distributed along a thick halo around the center of the shell, with higher concentration on the region corresponding to the fourth quadrant of the domain, with local variations resulting on a maximum error of approximately 26\%. The vertical displacement ${{u}}_z$ is also well reproduced, showing low RRMSE$_{\infty}$ across the domain with mean and median errors around 4\%. Local differences are more easily identifiable here, showing up on the regions located near the top-right corner, the red spot near the central hole and the very localised high displacement areas across the lateral boundaries, resulting on a maximum error of 17.4\%. For the remaining state variables the PB-PUNet model provides excellent results on approximating the global FEA solution with minimally noticeable local differences. For the stress resultants $\boldsymbol{{F}}$ and the membrane strains $\boldsymbol{\varepsilon}^m$ the model achieves mean and median errors less than or equal to 1\%. Higher errors are seen for the moments $\boldsymbol{{M}}$ and the flexural strains $\boldsymbol{\varepsilon}^f$, nevertheless, the mean and median are still within a very acceptable range of 2-4\%, approximately. %For both these state variables, local variations originate substantially higher maximum RRMSE$_\infty$ errors in the range 20-30\%, approximately.

%%%%%%%%%%%%%%%%%%%%%%%%%%%%%%%%%%%%
%% Results - GAN
%%%%%%%%%%%%%%%%%%%%%%%%%%%%%%%%%%%%
\subsection{AD-DCGAN}

\subsubsection{Latent manifold and membrane factor distribution}

\begin{figure}[!htbp]%
\centering
\includegraphics[width=\textwidth]{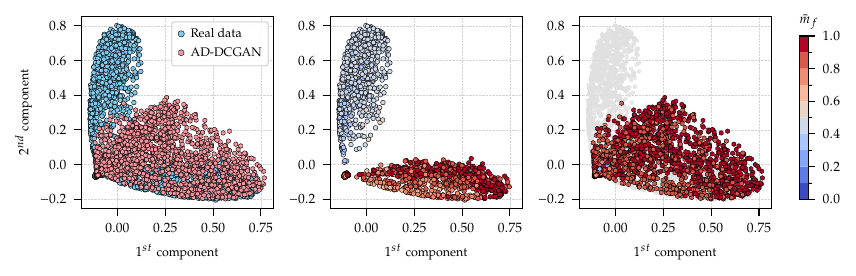}
{\caption{Features of the penultimate layer of the AD-DCGAN discriminator, corresponding to ground-truth and generated samples, projected onto a 2D latent manifold using kernel PCA algorithm.}
\label{fig:kpca}}
\end{figure}

The latent manifold representation of the features taken from the penultimate layer of the discriminator $D_{\text{shell}}$ is represented in Fig \ref{fig:kpca}. The first plot provides a global overview of the feature distributions of both real and generated samples over the latent manifold, while the second and third plots depict only the distributions corresponding to the real and generated samples, respectively. In both cases, the colour of each point in the latent space is associated with the average membrane factor $\bar{m_f}$ of the corresponding sample. 

\begin{figure}[!htbp]%
\centering
\includegraphics[width=0.425\textwidth]{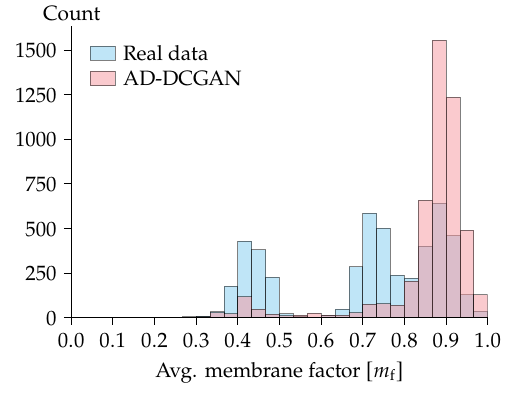}
{\caption{Average membrane factor distribution of the ground-truth and generated samples.}
\label{fig:hist_mf}}
\end{figure}

The results show that ground-truth features are densely packed across two petal-shaped clusters in the manifold, essentially grouping the training data into two categories: shells with low average membrane factor ($\bar{m}_{{f}}<0.6$ approx.) and shells with high membrane factors ($\bar{m}_{{f}}<0.6$ approx.). The features corresponding to generated samples are distributed towards the lower petal and along the space between the main clusters. Moreover, those features appear to be continuously spread out in space and do not show major signs of clustering, allowing to conclude the AD-DCGAN model does not suffer from mode collapse. By interpolating between both regions of the latent space the AD-DCGAN model shows a good capacity to generate new types of geometries, not seen in the training data. Additionally, the results clearly show the effect of the inductive bias introduced by the auxiliary discriminator $D_{\text{aux}}$ towards obtaining samples with higher average membrane factor. This is also supported by the histogram from Fig. \ref{fig:hist_mf}, which effectively shows the membrane factor distribution for generated samples skewed towards higher values, compared to the real data that encompasses a wider range of membrane factors.  

\subsubsection{Generated surfaces and FEA validation}

\begin{figure}[h!]%
\centering
\includegraphics[width=0.98\textwidth]{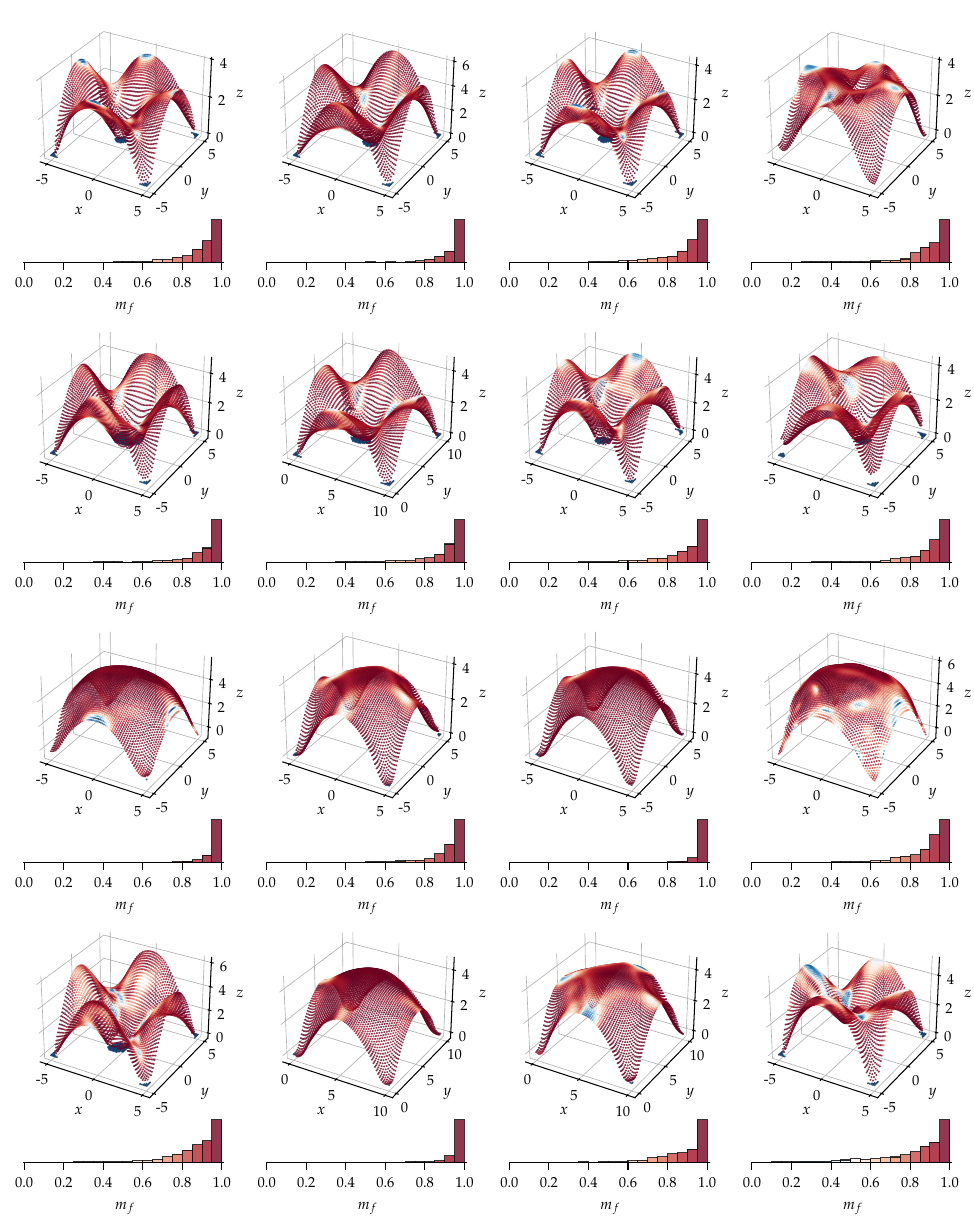}
{\caption{Post-processed surfaces generated by the AD-DCGAN and corresponding membrane factor distribution ${{m}}_{{f}}$.}
\label{fig:gen_smooth_shells}}
\end{figure}

The raw output obtained for a batch of 16 shell geometries generated by the AD-DDCGAN model is shown in \ref{ap:gen_surf_raw}. It can be seen that the model is capable of producing samples with clearly perceptible forms, albeit with a certain level of noise. However, FEA is a crucial step to validate these surfaces and requires a clean geometry in order to avoid meshing issues that may originate spurious solutions. As such, in an intermediate post-processing stage, the same batch of generated surfaces was passed through a radial basis function interpolator to obtain smooth surfaces. Keeping in mind that the smoothing slightly changes the geometries, the membrane factors were also recalculated. The post-processed surfaces are shown in Fig. \ref{fig:gen_smooth_shells} together with histograms that depict the corresponding membrane factor distributions. Observing these results, the histograms clearly demonstrate that the model shows a tendency to generate shell surfaces with high membrane factors, in line with the conclusions taken in the previous section. The AD-DCGAN is also capable of generating a variety of different geometries, encompassing both types of surfaces present in the training data (with and without central hole).

\begin{figure}[h!]%
\centering
\includegraphics[width=0.99\textwidth]{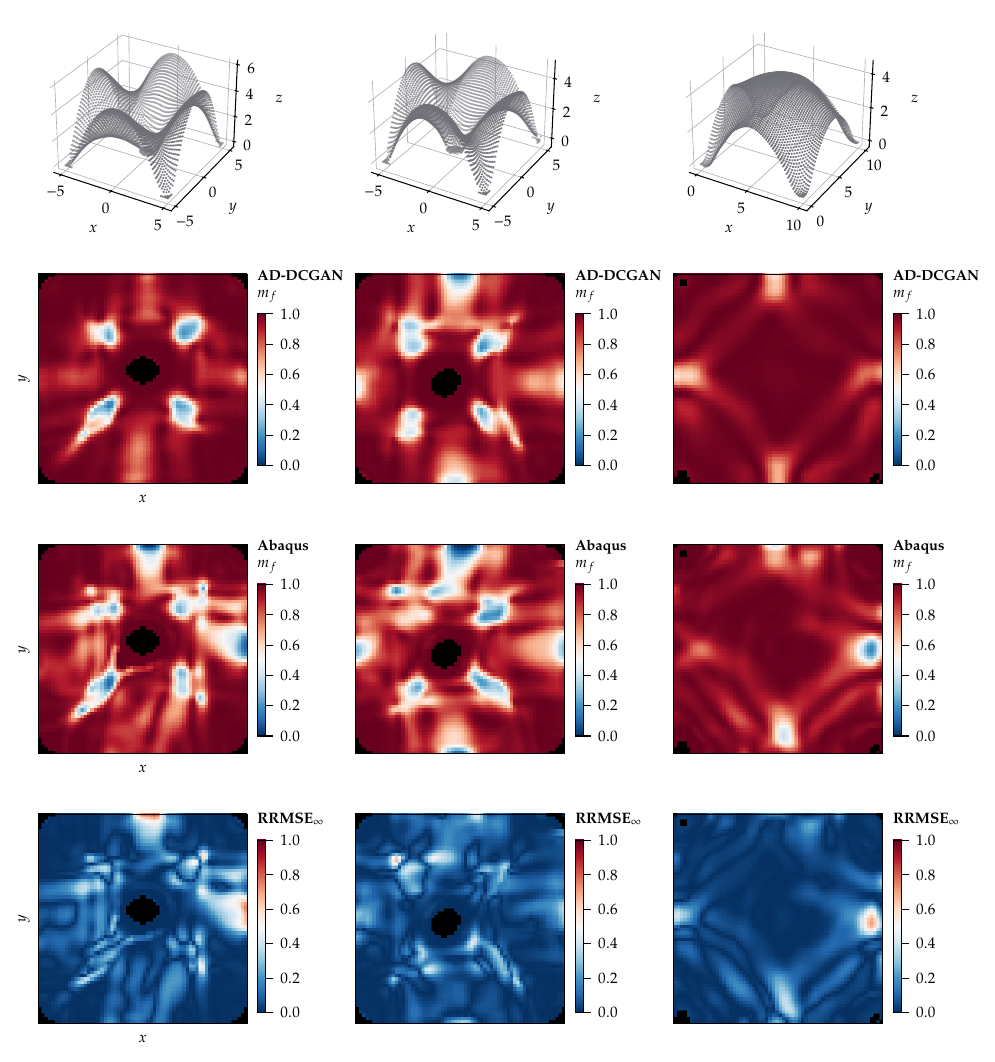}
{\caption{Pixel-wise $\text{RRMSE}_{\infty}$ prediction error for the membrane factor distributions ${{m}}_{{f}}$ of three surfaces generated by the AD-DCGAN model. Contour maps presented column-wise for each surface with corresponding 3D representation at the top.}
\label{fig:gen_shells}}
\end{figure}

From the batch of generated surfaces, three of them were chosen for validation using FEA. Here, the membrane factor distributions of the generated surfaces are compared with the corresponding FEA solution, which is taken as reference. The final results are presented in Fig. \ref{fig:gen_shells}. It can be seen that, in general, the membrane factors of the generated surfaces do not perfectly match the FEA solution, although the main features of the global behaviour are surprisingly well reproduced to a certain level. These differences could be attributed to several factors. First, the noisy output from the AD-DCGAN may cause the PB-PUNet to provide inaccurate predictions, given the fact that the latter is trained with data coming from smooth, clean surfaces. A possible solution would be to incorporate into the training process a post-processing step to smooth the generator output prior to feeding the data to the PB-PUNet, serving as auxiliary discriminator. A downside is that this would require the set of post-processing operations to be differentiable in order not to break the backpropagation. The results may also be hindered by the fact that the AD-DCGAN has the capacity to interpolate in the latent manifold, meaning that it may generate completely new surfaces with such a physical behaviour that may cause the PB-PUNet to extrapolate, which surrogate models struggle with, thus accumulating errors. In this sense, a physics-informed surrogate model based on a Graph Neural Network (GNN) architecture may provide superior results. Finally, the AD-DCGAN model merely generates new geometries and is not aware of the boundary conditions for each case, resulting in situations in which the surface extremities are not in total contact with the ground level, or it is not clear which nodes should be fixed. To mitigate that, the AD-DCGAN could eventually be provided with such information during training to learn how to generate surfaces with clear boundaries.

\clearpage
%%%%%%%%%%%%%%%%%%%%%%%%%%%%%%%%%%%%
%% Conclusions
%%%%%%%%%%%%%%%%%%%%%%%%%%%%%%%%%%%%
\section{Concluding remarks}\label{sc:conclusions}

This work introduced a physics-informed generative adversarial framework for the design of funicular (compression-only) shell structures. The approach integrates a modified DCGAN architecture with two key physical guidance mechanisms: a pre-trained PB-PUNet model that acts as an auxiliary discriminator to enforce physical compliance and structural efficiency through the evaluation of the membrane factor distribution, and a mask discriminator that enables the generation of geometries with complex features such as central holes.

The PB-PUNet was developed as a surrogate model to predict the structural response of shell geometries under self-weight. The systematic hyperparameter tuning process allowed to obtain a final model demonstrating excellent predictive accuracy, achieving a total error as low as 2.2\% on the test set, confirming the model's suitability for guiding the generative process.

The proposed AD-DCGAN was trained to generate level set-based representations of shell geometries. The use of a PCA-guided latent space and feature matching effectively mitigated mode collapse, as evidenced by the continuous distribution of generated samples across the latent manifold. The inductive bias introduced by the auxiliary discriminator successfully guided the generation towards geometries dominated by membrane behaviour, as shown by the high membrane factor distributions in the generated samples.  

%%%%%%%%%%%%%%%%%%%%%%%%%%%%%%%%%%%%
%% Disclaimer
%%%%%%%%%%%%%%%%%%%%%%%%%%%%%%%%%%%%
\section*{Disclaimer}

Views and opinions expressed are however those of the author(s) only and do not necessarily reflect those of the European Union or the European Research Council Executive Agency. Neither the European Union nor the granting authority can be held responsible for them.

%%%%%%%%%%%%%%%%%%%%%%%%%%%%%%%%%%%%
%% Authorship
%%%%%%%%%%%%%%%%%%%%%%%%%%%%%%%%%%%%
\section*{CRediT authorship contribution statement}

\textbf{Rúben Lourenço:} Conceptualization, Data curation, Formal analysis, Investigation, Methodology, Software, Validation, Visualization, Writing - original draft, Writing – review \& editing. \textbf{Icíar Alfaro:} Conceptualization, Data curation, Formal analysis, Investigation, Methodology, Software, Validation, Visualization, Writing – review \& editing. \textbf{Beatriz Moya:} Conceptualization, Formal analysis, Investigation, Methodology, Resources, Software, Supervision, Writing – review \& editing. \textbf{Elias Cueto:} Conceptualization, Formal analysis, Funding aquisition, Project administration, Resources, Supervision, Writing – review \& editing. All authors approved the final submitted draft.

%%%%%%%%%%%%%%%%%%%%%%%%%%%%%%%%%%%%
%% Competing interest
%%%%%%%%%%%%%%%%%%%%%%%%%%%%%%%%%%%%
\section*{Declaration of competing interest}

The authors declare that they have no known competing financial interests or personal relationships that could have appeared to influence the work reported in this paper.

%%%%%%%%%%%%%%%%%%%%%%%%%%%%%%%%%%%%
%% Data availability
%%%%%%%%%%%%%%%%%%%%%%%%%%%%%%%%%%%%
\section*{Data availability}

Data will be made available on request.

%%%%%%%%%%%%%%%%%%%%%%%%%%%%%%%%%%%%
%% Acknowledgements
%%%%%%%%%%%%%%%%%%%%%%%%%%%%%%%%%%%%
\section*{Acknowledgements}

The authors acknowledge the support of the Spanish Ministry of Science and Innovation, through AEI/10.13039/501100011033 (grant no. PID2023-147373OB-I00). The authors also acknowledge the support of the Ministry for Digital Transformation and the Civil Service, through the ENIA 2022 Chairs for the creation of the university-industry chairs in Artificial Intelligence, through grant TSI-100930-2023-1, and Keysight-ESI Group support through the ENIA Chair. 

This work is supported by ERC grant PHYSIA 101264273.

Beatriz Moya acknowledges the support of the French National Research Agency under the Junior Professor Chair contract ANR-24-CPJ1-0005-01.

\clearpage

%%%%%%%%%%%%%%%%%%%%%%%%%%%%%%%%%%%%
%% Appendix A
%%%%%%%%%%%%%%%%%%%%%%%%%%%%%%%%%%%%
\appendix
\section{PB-PUNet predicted state variables}\label{ap:pinn}

\begin{figure}[!htbp]%
\centering
\includegraphics[width=\textwidth]{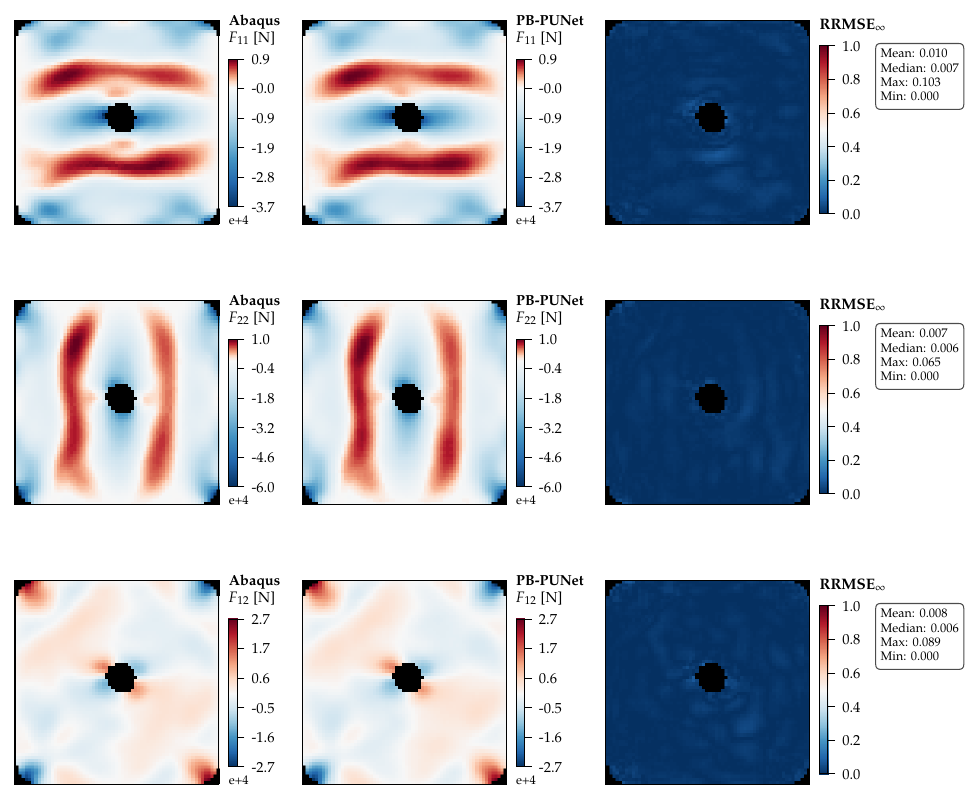}
{\caption{Pixel-wise RRMSE$_{\infty}$ prediction error of the PB-PUNet on the components of the internal force tensor $\boldsymbol{{F}}$ for the first sample of the test set.}
\label{fig:force}}
\end{figure}

\begin{figure}[!htbp]%
\centering
\includegraphics[width=\textwidth]{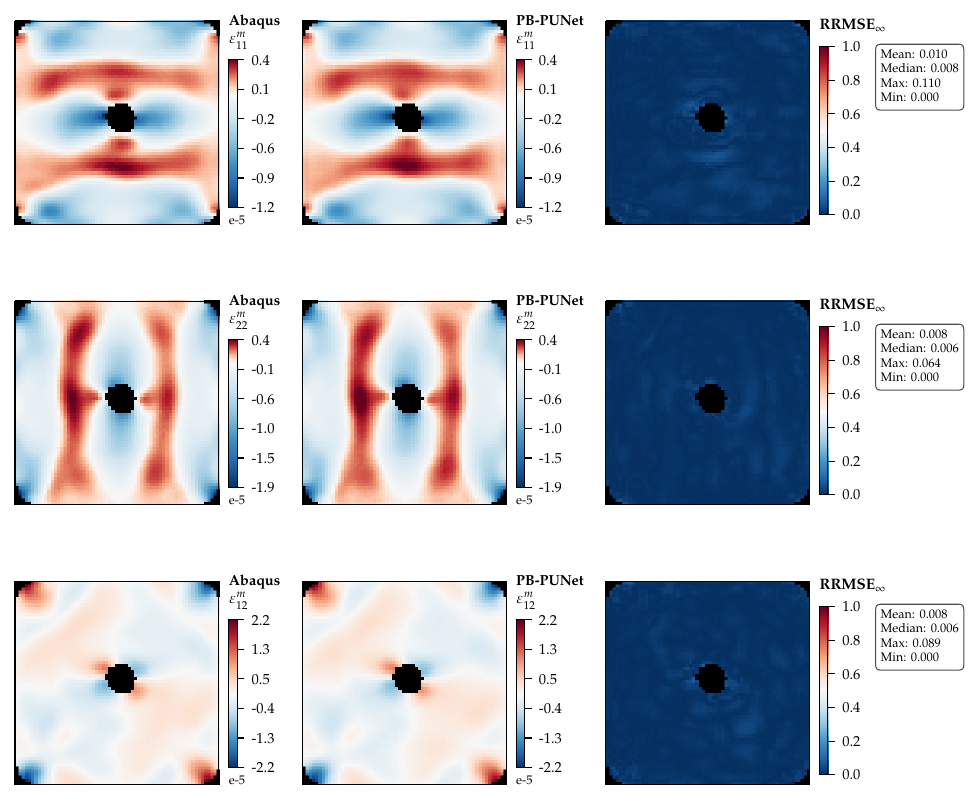}
{\caption{Pixel-wise RRMSE$_{\infty}$ prediction error of the PB-PUNet on the components of the membrane strain tensor $\boldsymbol{\varepsilon}^m$ for the first sample of the test set.}
\label{fig:strain}}
\end{figure}

\begin{figure}[!htbp]%
\centering
\includegraphics[width=\textwidth]{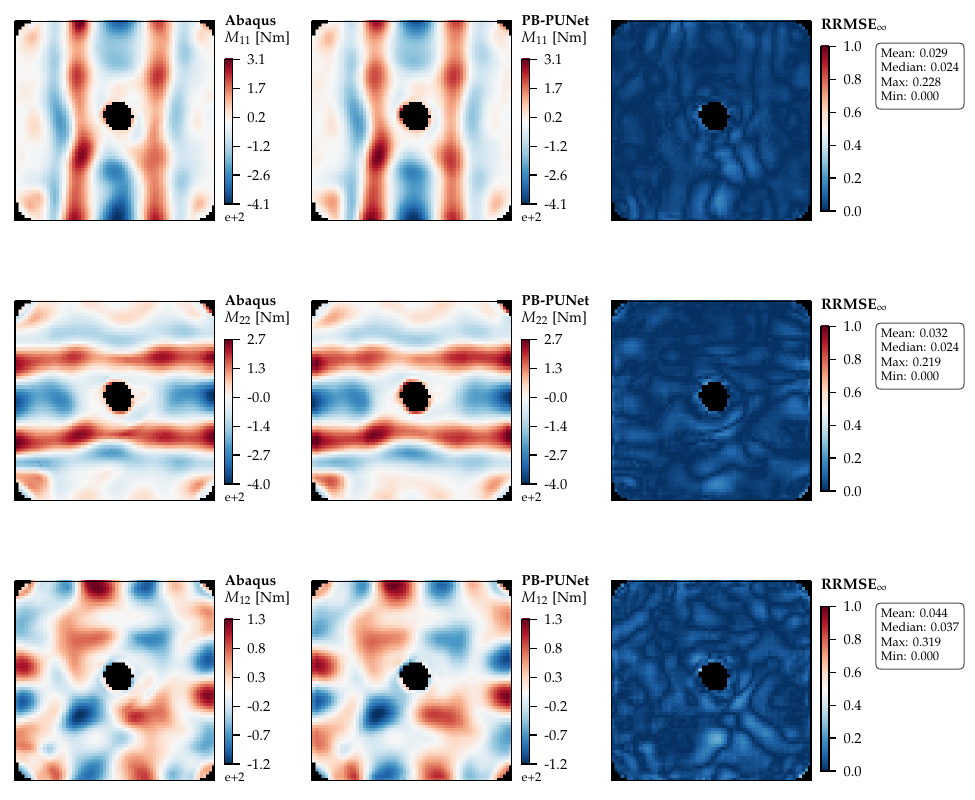}
{\caption{Pixel-wise RRMSE$_{\infty}$ prediction error of the PB-PUNet on the components of the bending moment tensor $\boldsymbol{{M}}$ for the first sample of the test set.}
\label{fig:moment}}
\end{figure}

\begin{figure}[!htbp]%
\centering
\includegraphics[width=\textwidth]{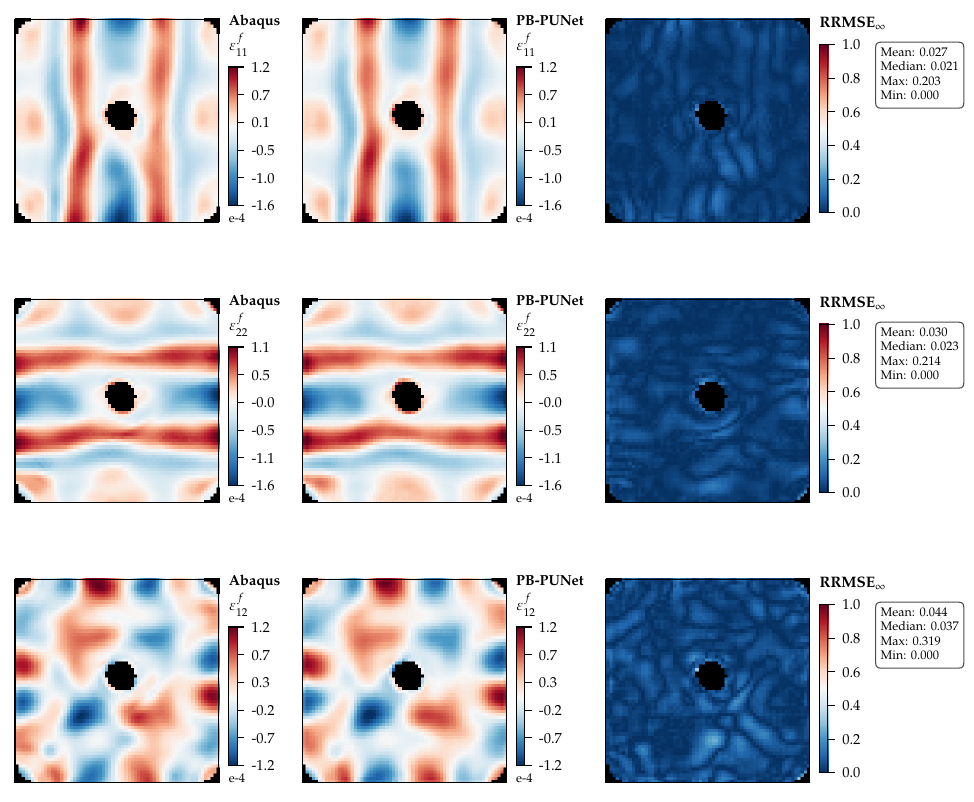}
{\caption{Pixel-wise RRMSE$_{\infty}$ prediction error of the PB-PUNet on the components of the flexural strain tensor $\boldsymbol{\varepsilon}^f$ for the first sample of the test set.}
\label{fig:twists}}
\end{figure}

\clearpage

%%%%%%%%%%%%%%%%%%%%%%%%%%%%%%%%%%%%
%% Appendix B
\section{AD-DCGAN generated surfaces}\label{ap:gen_surf_raw}

\begin{figure}[!htbp]%
\centering
\includegraphics[width=0.96\textwidth]
{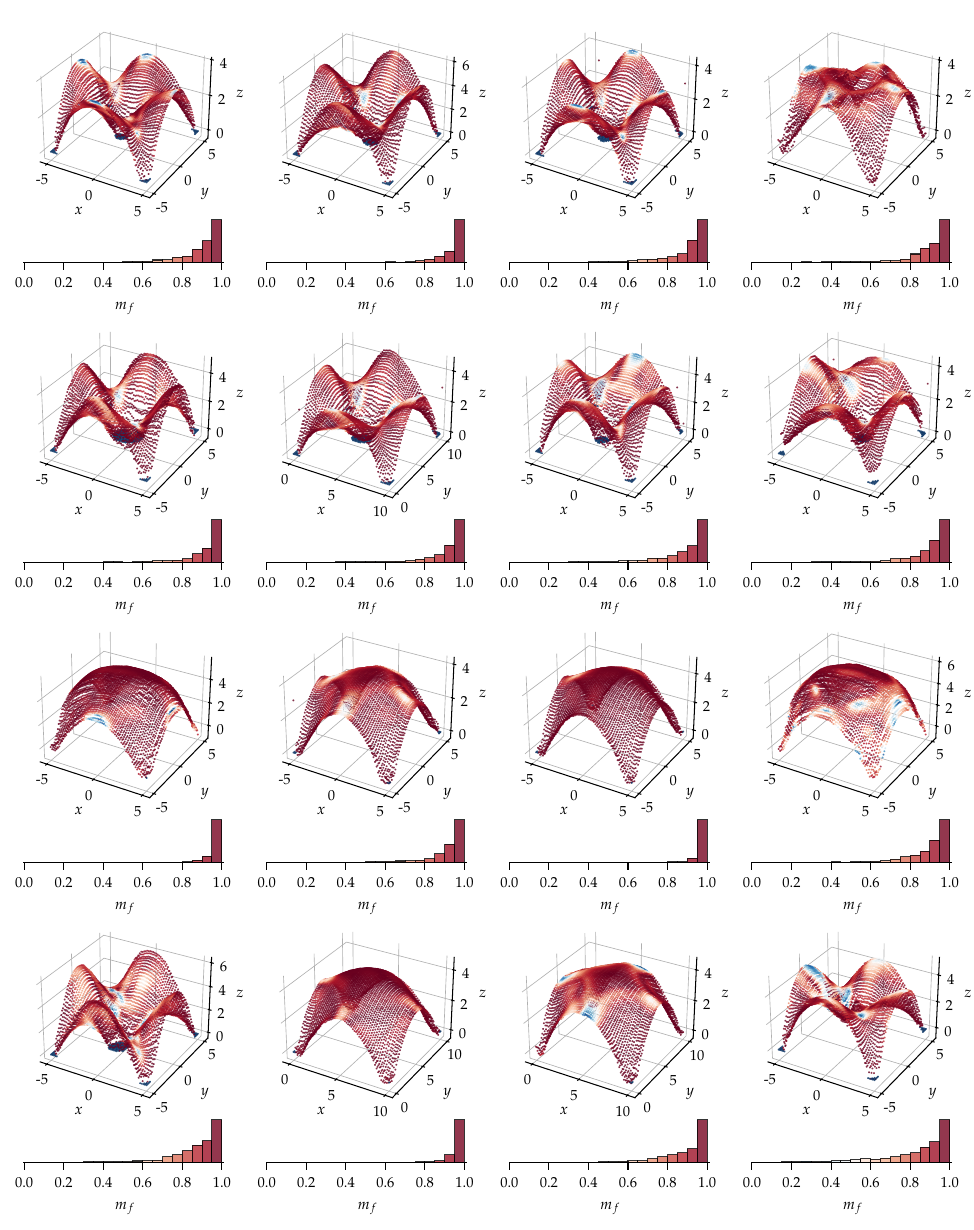}
\vspace{-3mm}
{\caption{Surfaces generated by the AD-DCGAN and corresponding membrane factor distribution ${{m}}_{{f}}$.}
\label{fig:gen_surf_raw}}
\end{figure}

\clearpage
%%%%%%%%%%%%%%%%%%%%%%%%%%%%%%%%%%%%

%%%%%%%%%%%%%%%%%%%%%%%%%%%%%%%%%%%%
%% Bibliography
%%%%%%%%%%%%%%%%%%%%%%%%%%%%%%%%%%%%
\bibliographystyle{elsarticle-num} 
\bibliography{refs.bib}
\end{document}